\documentclass[journal]{IEEEtran}
\usepackage{graphicx}
\usepackage{subfigure}  % support sub-figure
\usepackage[ruled]{algorithm2e}
\usepackage{multirow} %multirow for format of table 
\usepackage{amsmath} 
\usepackage{amssymb}
\usepackage{cite}
\usepackage{booktabs}
\usepackage{footnote}
\usepackage{makecell}
\usepackage{float}
\usepackage{url}

\ifCLASSINFOpdf
\else
\fi
\begin{document}
% \title{When Digital Twin Meets Intelligent Transportation Systems: A Federated Learning Solution}
\title{Density-Aware Federated Imitation Learning for Connected and Automated Vehicles with Unsignalized Intersection}

% \author{Michael~Shell,~\IEEEmembership{Member,~IEEE,}
%         John~Doe,~\IEEEmembership{Fellow,~OSA,}
%         and~Jane~Doe,~\IEEEmembership{Life~Fellow,~IEEE}% <-this % stops a space
\author{Tianhao~Wu,
  Mingzhi~Jiang,
  Yinhui~Han,
  Zheng~Yuan,
  % Zhe~Wang,
  % Xinhang~Li,
  Lin~Zhang% <-this % stops a space
  \thanks{T. Wu, M. Jiang, Y. Han and Z. Yuan are with the School of Artificial Intelligence, Beijing University of Posts and Telecommunications, Beijing, 100876 China.}% <-this % stops a space
  \thanks{L. Zhang is with Beijing Information Science and Technology University, Beijing 100192 China. He is also with Beijing University of Posts and Telecommunications, Beijing 100876 China e-mail:zhl@bistu.edu.cn.}
  % \thanks{J. Doe and J. Doe are with Anonymous University.}% <-this % stops a space
  \thanks{Manuscript received *** **, ****; revised *** **, ****.}}


\markboth{IEEE Internet of Things Journal,~Vol.~**, No.~*, ***~****}%
{Tianhao \MakeLowercase{\textit{et al.}}: Density-Aware Federated Imitation Learning for Connected and Automated Vehicles with Unsignalized Intersection}



\maketitle

\begin{abstract}

  Intelligent Transportation System (ITS) has become one of the essential components in Industry 4.0. As one of the critical indicators of ITS, efficiency has attracted wide attention from researchers.
  However, the next generation of urban traffic carried by multiple transport service providers may prohibit the raw data interaction among multiple regions for privacy reasons, easily ignored in the existing research.
  This paper puts forward a federated learning-based vehicle control framework to solve the above problem, including interactors, trainers, and an aggregator. In addition, the density-aware model aggregation method is utilized in this framework to improve vehicle control.
  What is more, to promote the performance of the end-to-end learning algorithm in the safety aspect, this paper proposes an imitation learning algorithm, which can obtain collision avoidance capabilities from a set of collision avoidance rules.
  Furthermore, a loss-aware experience selection strategy is also explored, reducing the communication overhead between the interactors and the trainers via extra computing.
  Finally, the experiment results demonstrate that the proposed imitation learning algorithm obtains the ability to avoid collisions and reduces discomfort by 55.71\%. Besides, density-aware model aggregation can further reduce discomfort by 41.37\%, and the experience selection scheme can reduce the communication overhead by 12.80\% while ensuring model convergence.

\end{abstract}


% \fi
\IEEEpeerreviewmaketitle

\section{Introduction}

%%%%%%%%% ITS %%%%%%%%%
\IEEEPARstart{T}he development of Industry 4.0 has profoundly promoted the transformation of the manufacturing industry to intelligence and automation, leading to the integration of information network technology and industry.
Meanwhile, as a critical part of Industry 4.0, the manufacturing industry also presents a tendency of intelligence and high integration.
More importantly, industry 4.0 allows manufacturing factories distributed in different regions to flexibly dispatch resources, which puts forward higher demand for transportation\cite{9068495}.
According to related research\cite{9363013}, Intelligent Transportation System (ITS) has an advantage in improving safety, efficiency and driving comfort.
Connected and Automated Vehicles (CAVs)\cite{wang2019survey} is an indispensable enabler for ITS, derived from advanced communication technology and emerging computing technology.
CAVs are expected to revolutionize future ITS to solve the issues of safety\cite{rios2016automated}, efficiency\cite{fernandes2014multiplatooning}, and sustainability\cite{altan2017glidepath}.
With CAVs, traffic optimization can occur at different places,  including automated intersection management\cite{DBLP:journals/tcps/KhayatianMADCLS20}.

%%%%%%%%% intersection %%%%%%%%%
Compared to conventional traffic light control, automated intersection management aims to achieve higher throughput while ensuring the safety of vehicles.
The process of deciding the vehicle passing sequence is called scheduling.
The three main categories of existing scheduling policies are:
1) First-Come First-Served\cite{DBLP:journals/jair/DresnerS08}: The first arrival vehicle is to be served firstly and entry to the intersection.
2) Optimization-Based\cite{DBLP:journals/tits/BichiouR19}: The approaches try to minimize the average travel time for all vehicles passing the intersection, and the passing sequence may vary from the original.
3) Heuristic\cite{DBLP:conf/itsc/StevanovicM18}: The optimal solution is not guaranteed, but it is sufficient to reach the immediate goal.
Because the actions taken by CAVs depend on the real-time driving conditions, which is a typical Markov Decision Process (MDP), Reinforcement Learning (RL) is suitable to address this issue.
Guan et al. proposed an RL-based method to centralized guide a fixed number of vehicles through the intersection \cite{guan2020centralized}.
Wu et al. decoupled the relationship between the identity and driving information of vehicles and proposed a cooperative RL method to improve traffic efficiency while ensuring safety\cite{wu2020cooperative}.
Jiang et al. proposed a two-stage RL incorporating end-edge-cloud architecture to achieve global optimization among multiple homogeneous intersections \cite{jiang2020multi}.
However, low sample efficiency and limited safety performance bring a great difficulty to practical application.

%%%%%%%%% IL %%%%%%%%%

%%%%%%%%% FL %%%%%%%%%

Although some progress has been achieved in learning-based vehicle control at intersections, three problems still need to be solved at different intersections:
1) Isolation: For better vehicle control performance, it is essential to set a local center to assist the vehicle control, which means that some privacy-sensitive data is kept in a local center. Due to the privacy requirements of transport service providers, the data will not be transmitted to cloud nodes or other peer nodes.
2) Heterogeneity: Due to different traffic densities at intersections, the generated experience data drive the obtained RL model to show different capabilities for vehicle control optimization. Therefore, the conventional model parameter averaging may not meet the performance requirements at different intersections.
3) Scalability: As the number of CAV-enabled unsignalized intersections grows, data generated by the vehicles increases. Any learning-based algorithm with a centralized property may find it difficult to handle such data due to the incurred high computation and communication budget.

\begin{figure*}
  \centering
  \includegraphics[width=\textwidth]{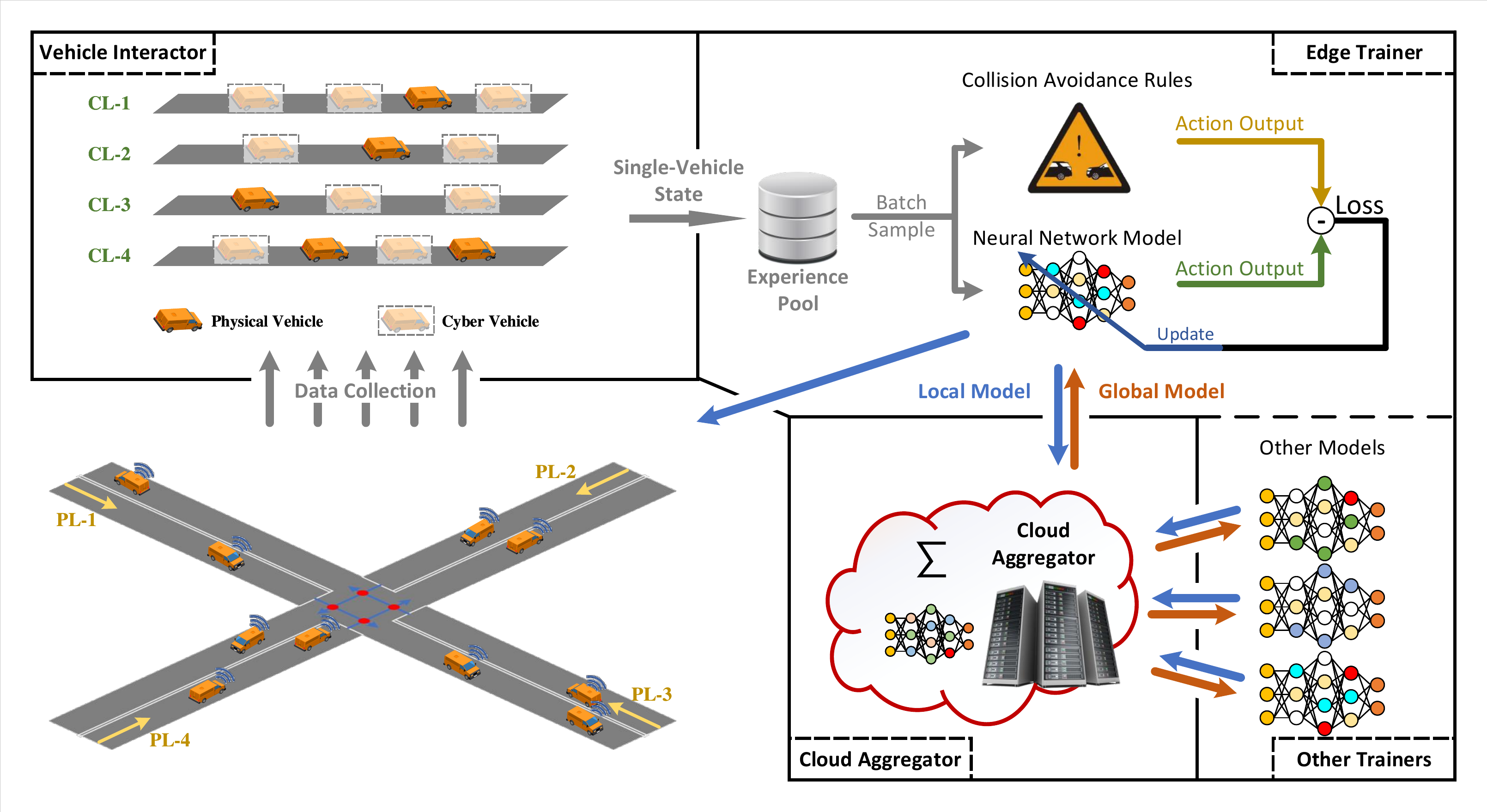}
  \caption{Federated Imitation Learning for Distributed Vehicle Control. Vehicles take the role of interactors, collect the surrounding information and map other vehicles from Physical-Lanes (PLs) to Cyber-Lanes (CLs). The reorganized information is upload to the experience pool as a piece of experience. The edge trainers sample the experience to train a local model via the proposed imitation learning algorithm. Then, several local models are collected at cloud nodes to be aggregated as a global model. Next, the global model is distributed to the edge trainer for subsequent updates. After several iterations, the final vehicle control model is obtained.}
  \label{fig:General} %% label for entire figure
\end{figure*}

To cope with the above-mentioned challenges, a federated deep learning framework, as shown in Fig.\ref{fig:General}, is proposed for vehicle control policy acquisition at various intersections.
The main contributions of this article are summarized as follows.
\begin{itemize}
  \item This paper proposes a density-aware federated learning (FL) framework for unsignalized intersection control, reducing communication costs and supporting various vehicle densities. This framework includes four main components: data collection, experience upload, model training and model aggregation.
  \item An Imitation Learning (IL) algorithm is proposed to obtain a safety-oriented vehicle control policy, which trains the model with the experience from collision avoidance rules. Compared with the pure rule strategy, the model training process makes up for the lack of driving comfort.
  \item In response to limited communication, a loss-aware experience selection strategy is designed, which can reduce communication overhead by extra computation. Trainers output and deliver a reference loss as a threshold. Each interactor compares the newly-generated experience with the threshold to decide whether to upload.
  % generates new experiences and decide whether to upload based on the threshold.
  \item The extensive experiment is conducted, and the results demonstrate that IL can significantly improve safety and reduce discomfort by 55.71\%, FL combined with IL can further reduce discomfort by 41.37\%, and experience selection strategy can reduce the communication overhead by up to 12.80\%.
\end{itemize}

The rest of this article is organized as follows:
The system architecture is present in Section \ref{sec:system}.
Section \ref{sec:FIL} elaborates the details of the proposed federated deep learning framework and IL algorithm for vehicle control.
In Section \ref{sec:loss-aware}, a loss-aware experience selection strategy is presented.
The performance evaluation and analysis are provided in Section \ref{sec:experiment}.
Section \ref{sec:conclusion} concludes this article.

\section{System Architecture}
\label{sec:system}

This section mainly considers a hierarchical network in an urban scenario, which consists of one Cloud Aggregator (CA), many Edge Trainers (ETs) and lots of Vehicle Interactors (VIs), as shown in Fig.\ref{fig:General}.
The CA is deployed in the remote cloud and connects to a set of ETs via a reliable backhaul link. These ETs are denoted by $\mathbb{E} = \{E_{1}, E_{2}, \dots, E_{n}, \dots, E_{N}\}$, where $N$ is the number of ETs.
Each ET $i$ serves its wirelessly connected VIs, denoted by $ \mathbb{V}_{n} = \{V_{1}, V_{2}, \dots, V_{i}, \dots, V_{m(t)}\}$. Note that, under an ET, the number of VI, $m(t)$, varies over time $t$.
Both ETs and VIs are equipped with powerful GPU computing services.
The ETs are used to train the local model by the information uploaded by connected VIs. The CA is used to produce a global model by combining the local information of different ETs.

On the aspect of unsignalized intersection control, there are three assumptions to support our work, which is similar to related work\cite{bian2019cooperation}.
\begin{itemize}
  \item Longitudinal vehicle control is the focus of this paper. All vehicles keep their original directions and go straight within the intersection area.
  \item All vehicles can measure kinetic information, strictly obey the determined acceleration and communicate with adjacent nodes, i.e. edge trains and adjacent vehicles.
  \item Communication latency and package loss are not taken into consideration for simplification.
\end{itemize}

The longitudinal motion of vehicles is given by
\begin{equation}
  \begin{aligned}
    x^{long}(t + 1) & = x^{long}(t) - v(t)T  - \frac{1}{2}a(t)T^2 \\
    v(t + 1)        & = v(t) + a(t)T
  \end{aligned},
  \label{eq:x_v_t}
\end{equation}
where $x^{long}$ is the displacement, $v$ and $a$ are the velocity and acceleration respectively, and $T$ is the discrete-time step. The change of vehicle motion state depends on the input, i.e. acceleration, at the previous time step.

This paper adopts distributed decision-making for vehicle control. That is to say, each vehicle constructs its cyberspace and maps adjacent vehicles to cyber objects.
As a result, vehicle $j$ decides its action $a_j$ based on its surroundings.
\begin{equation}
  {a_i} = P(\overrightarrow {s_i} |\theta )
\end{equation}
where $P(\cdot|\theta)$ is $\theta$-parameterized policy for decision-making. $\overrightarrow {s_{i}}$ is the state vector of the vehicle $i$ and $\overrightarrow {s_{i}} = {s_i},\overrightarrow {s_{ - i}}$.
$s_i$ is state of the ego-vehicle $i$, including position, velocity and acceleration. Moreover, $\overrightarrow {{s_{ - i}}}$ is a vehicle set other than vehicle $i$. $|\overrightarrow {{s_{ - i}}}|$, the number of vehicle set $\overrightarrow {{s_{ - i}}}$, is defined by a selection scheme.
To simulate traffic flow, the number of arrived vehicles for entering each intersection during the period $t$ is defined as $V_{q}(t)$. It follows a Poisson process with the parameter $\lambda$:
\begin{equation}
  P\left(V_{q}(t)=g\right)=\frac{(\lambda t)^{g}}{g !} e^{-\lambda t}
\end{equation}
where $g$ equals the number of vehicles generated in a period $t$.
The introduction of the Poisson process means that vehicles are created dynamically, which is similar to real traffic.

As shown in Fig.\ref{fig:General}, it is a three-layered architecture.
The bottom layer contains VIs requesting model downloading.
The middle layer includes several ETs equipped with GPU computing servers and experience databases.
The top layer has a global model aggregator.
The multiple connected VIs interact with the environment and individually upload experience data to the local database.
Each EI firstly generates a local model and acquires the experience from connected VIs. Then, based on the received data, each ET utilizes the local GPU computing ability to compute a local model. Next, the ET sends the local model to VIs for vehicle control and the CA for global model aggregation. Finally, the CA aggregates the models and sends the global model back to each ET.
The above steps are repeated until a satisfying global model is achieved.
The trained model in this work is specially developed to output the vehicles' accelerations in response to the contextual information around the vehicle.

\section{Density-aware federated imitation learning for vehicle control}
\label{sec:FIL}

This section elaborates on the proposed vehicle control scheme.
Firstly, the density-aware federated deep learning framework is described, including model download, experience upload, FL model training, upload updated model and weighted aggregation.
Then, we introduce a set of collision avoidance rules as a basis for further optimization.
Finally, with the rule, an IL algorithm is proposed for vehicle control.

\subsection{Density-aware Federated Deep Learning Framework}

FL enables collaborative training of a deep neural network model among ETs under the orchestration of a server in CA by keeping the training data on each ET at intersections.
And it not only significantly reduces the privacy risk of the vehicle but also dramatically reduces the communication overhead caused by the centralized machine learning \cite{DBLP:journals/corr/KonecnyMYRSB16}.
FL is enabled by multiple communication rounds (computing iteration).
$N$ intersections with known traffic densities and the corresponding ETs are selected to conduct model training.
The $N$ ETs are indexed by $n$.
Then, each ETs retrieves a global model from the CA and trains this model from its local model with the data collected from vehicles in the intersection area.
Following the local training of ETs, the updated weights and gradients are sent back to the CA. Finally, the CA aggregates the collected models from vehicles to construct an updated global model.
Furthermore, the final-trained models are distributed to the vehicles to ensure that vehicles pass through the intersection.

The details of the proposed designed FL iteration consists of the following steps:
\subsubsection{Model Download}

A set of intersections are chosen to take part in FL training.
The ETs at these intersections download the global model $\omega_r $ from the CA and train this model over their local data.

\subsubsection{Experience Upload}

To achieve better vehicle control performance, each vehicle needs to consider other vehicles' states for inference. However, due to the insufficient number of collected samples and non-uniform distribution, the efficiency of vehicular distributed training is very low.
Under the above setting, it is essential to adopt centralized training and distributed execution\cite{DBLP:journals/corr/KonecnyMYRSB16}.
Then each vehicle interacts with the environment, i.e. other vehicles, and generates enormous experience data to upload.
This data is used to train the local model by the corresponding ET.

\subsubsection{FL Model Training}

The third step in the proposed FL is to train the model by utilizing local data uploaded by vehicles.
Let $Exp = \{Exp_{1}, Exp_{2}, \dots, Exp_{n}, \dots, Exp_{N}\}$ represent the experience data stored in selected ETs.
$Exp_{n}$ denotes the local experience of the $n^{th}$ ET with the length $d_n$, $d_n=|Exp_{n}|$. $d$ is the size of the whole data among the selected ETs.
The goal of the FL is to minimize the loss function $L(\omega)$:
\begin{equation}
  \begin{aligned}
    \min _{\omega} L(\omega) & =\sum_{n=1}^{N} \frac{d_{n}}{d} L_{n}(\omega) \quad \text { where } \\
    L_{n}(\omega)            & =\frac{1}{d_{j}} \sum_{j \in H_{k}} l_{j}(\omega),
  \end{aligned}
\end{equation}
where $l_{j}(\omega)$ is the loss of the vehicle control on the $j^{th}$ experience batch in $Exp_n$ with the parameters of model $\omega$, and $d_{j}$ is the experience number in the $j^{th}$ batch. $n$ denotes the index of total selected ETs $N$.
$L_{n}(\omega)$ represents the local loss function of ET $n$.
Then it is obviously that minimizing the weighted average of local loss function $L_{n}(\omega)$ is equal to optimize the loss function $l(\omega)$ of FL.

\subsubsection{Upload Updated Model}

The fourth step is to upload the local model ${\omega}^{n}_{t+1}$ from ETs to the CA.
The communication cost exceeds the computing cost\cite{DBLP:conf/aistats/McMahanMRHA17}.
In order to reduce communication overhead, the model can be compressed before being uploaded to the CA.

\subsubsection{Weighted Aggregation}

After ETs uploading their models, the fifth step is to produce the new global model $\omega_{r+1}$ by computing a weighted sum of all received models ${\omega}^{n}_{r}$.
The new generated global model is used for the next training iteration. $r$ denotes the communication rounds in FL.
Compared with the typical Federated Stochastic Gradient Descent (FedSGD), Federated Averaging (FedAVG), widely applied in FL, increases the proportion of local computing and decreases mini-batch sizes.
In FedSGD, each ET $n$ utilizes its local data to locally compute the average gradient $\nabla L_n(\omega)$ on its global model $omega_r$.
The CA then aggregates these computed gradients by taking a weighted average sum and applies update gradients:
\begin{equation}
  w_{r+1} \leftarrow w_{r}-\eta \sum_{n=1}^{N} \frac{d_{n}}{d} w_{r+1}^{n},
\end{equation}
where $\eta$ is the static learning rate. Whereas, in FedAVG, each ET adds more computation by iterating the local updates $ w_{r}^{n} \leftarrow w_{r}^{n}-\eta \nabla L_{n}\left(w_{r}^{n}\right)$ multiple times before the aggregation step in the CA.
The weighted averaging algorithm is implemented to aggregate the model.
Weights for parameter aggregation is dependent on the traffic density of each intersection, which is $\gamma_{n} = \mathfrak{D}_n / \mathfrak{D}$. $\mathfrak{D}_n$ and $\mathfrak{D}$ respectively denote the traffic density on $n^{th}$ intersections and the density sum of all intersections.
Then, the aggregate method can be re-written as
\begin{equation}
  w_{r+1} \leftarrow w_{r}-\eta \sum_{n=1}^{N} \gamma_n \frac{d_{n}}{d} w_{r+1}^{n}.
\end{equation}
Selected intersections with a higher traffic density account for more contributions and are given greater weight in model aggregation.

\subsection{Imitation Learning for vehicle control}

The model trained in the above FL framework is the IL model for vehicle control.
It is used to obtain the vehicle control capability from the collision avoidance rules.
Both IL and RL depend on environment interaction. Unlike RL, which obtains the desired behaviors according to the hidden objectives, IL directly clones the desired behaviors.
IL can overcome the highly uncertain initial state and the sparse reward, which exists in RL and may lead to an exploration trap.
Thus, this part explains how to imitate the end-to-end vehicle control policy from existing rules.

As presented in Fig.\ref{fig:General}, there are two modules to support vehicle control policy acquisition. The first module is a set of collision avoidance rules to output expert experience, described in Section \ref{sec:rule}. The second module is a continuous updated deep neural network as the final vehicle control policy carrier.
The proposed IL consists of the following steps:
\begin{enumerate}
  \item Collision avoidance rules guide vehicles to make actions at intersections. The state $s$ and $a$ are recorded as expert experience $(s,a)$ to upload to the corresponding ET's experience buffer.
  \item ET sample a batch of experience from the pool for the process of IL.
  \item Experience batch is simultaneously forwarded into two modules, collision avoidance rules and deep neural network.
  \item The loss is calculated with the output of the two modules. This paper utilizes the square of the difference between the two outputs to calculate the loss.
  \item The deep neural network is updated by minimizing the loss mentioned above.
\end{enumerate}
The loss function of IL model update can be defined as below,
\begin{equation}
  \label{eq:IL_loss}
  L(\overrightarrow {Exp}){\rm{ = }}\frac{1}{B}\sum\limits_{i = 1}^B {{{\left| {{\pi _\theta }\left( {{{s}_i}} \right) - {a_i}} \right|}^2}} ,
\end{equation}
where $\overrightarrow {Exp}$ is a batch of experience, and the batch size is $B$, given in the experiment part.
$\pi _\theta(\cdot)$ is $\theta$-parameterized policy, and each iteration updates the parameter $\theta$.
With collision avoidance rules integrated, the loss function can drive the deep neural network to produce safety-oriented strategies.

\subsection{Collision Avoidance Rules}
\label{sec:rule}
In this part, a concept of Cyber-Lane (CL) is introduced to reconstruct vehicles' relationship.
Vehicles in different trajectories travel in different Physical-Lanes (PLs), and these trajectories intersect into conflict points.
All vehicles are projected on the CLs from PLs based on conflict points.
From the perspective of CL, the position relationship of all vehicles is reorganized.
After being projected, vehicle $A1$ appears between vehicle $B1$ and $B2$. Then, the action of vehicle $B1$ and $B2$ will naturally take into account the action of vehicle $A1$.

The set of rules considers three factors, including space, time and acceleration.
Space is the first factor, which can directly determine whether the collision has occurred.
As a second-order factor, time considers whether the vehicle will collide in the near future.
The acceleration indicates whether the collision will be avoided.
In this paper, the above factors are quantified as safety values (SV).

The first factor is space. The safety value for space $SV_{j,s}$ is calculated as below,
% alpha_{s} 10
% beta_{s}10
\begin{equation}
  SV_{j,s}=log((\frac{d_{j,nearest}}{\alpha_{s}})^{\beta_{s}}),
  \label{eq:d_safe_value}
\end{equation}
where $d_{j,nearest}$ denotes the distance between vehicle $j$ and its nearest vehicle on the virtual lane. $\alpha_{s}$ normalizes $d_{j,nearest}$, and it can be treated as the expected headway distance. $\beta_{s}$ increases the offset to improve $log(\cdot)$ effect.
There is a positive correlation between safety value and nearest inter-vehicle space distance.

The second factor is time. The SV for time $SV_{j,t}$ is calculated as below,
% alpha_t 1.5
% beta_t 2
\begin{equation}
  {SV_{j,t}} = \left\{ {\begin{array}{*{20}{c}}
        { - {{\left[ {\frac{{\alpha_{t}}}{{\tanh ( - {t_{j,nearest}})}}} \right]}^{\beta_t}}} & {0 < {t_{j,nearest}} < 1} \\
        2                                                                                     & {otherwise}
      \end{array}}, \right.
  \label{eq:t_safe_value}
\end{equation}
where $t_{i,nearest}$ denotes Time To Collision (TTC) between vehicle $j$ and its nearest vehicle. In the sensitive range, where $t_{i,nearest}$ is no more than 1, the function $tanh(\cdot)$ is used to mark the nearby collision risk.
There is a negative correlation between safety value and TTC.

The third factor is acceleration. The SV for acceleration $SV_{j,acc}$ is calculated as below,
\begin{equation}
  SV_{j,acc} = \lambda_{acc} \times acc_{j,front} \times \log \left( {\min {{\left( {\frac{{{d_{j,front}}}}{{{d_{threshold}}}},\alpha_{acc}} \right)}^{\beta_{acc}}}} \right),
  \label{eq:acc_safe_value}
\end{equation}
where $d_{j,front}$ is the distance from vehicle $j$ to its front vehicle, $acc_{j,f}$ is the acceleration of the vehicle in front of vehicle $j$, and $d_{threshold}$ is the space distance safety threshold.
$min(\cdot)$ is used to control the range of $\frac{{{d_{j,front}}}}{{{d_{threshold}}}}$ within $[0,\alpha_{acc} ]$.
To limit the influence of acceleration in the calculation of safety value, discount factor $\lambda_{acc}$ is introduced.

The combination of SV is calculated as follows,
% 最大值 SV_{max} 20
% 最小值 SV_{min} -20
% lambda 0.2
\begin{equation}
  \begin{array}{*{5}{l}}
    S{V_j} & = Comb\left( {S{V_{j,s}},S{V_{j,t}},S{V_{j,acc}} | S{V_{max}},S{V_{min}}} \right)                     \\
           & = clip\left( {\left( {S{V_{j,d}} + S{V_{j,t}} + S{V_{j,acc}}} \right),S{V_{max}},S{V_{min}}} \right),
  \end{array}
  \label{eq:safe_value_combination}
\end{equation}
where $SV_{j,d}$, $SV_{j,t}$, and$SV_{j,acc}$ are defined above. In order to obtain a proper acceleration value in Eq.\ref{eq:safe_to_acc}, $clip(\cdot)$ is used to limit the maximum and minimum.
A larger $S{V_j}$ indicates the vehicle $j$ driving in a safer condition.
Based on the above SV, the ego vehicle's action can be calculated as follows,
\begin{equation}
  a_{exe} = \left\{ {\begin{array}{*{20}{c}}
        {\left| {\frac{{SV}}{\eta}} \right|} & {d_f \leq d_b} \\
        {\frac{{SV}}{\eta}}                  & {d_f > d_b}
      \end{array}} \right.,
  \label{eq:safe_to_acc}
\end{equation}
where $d_f$ is the distance to the vehicle in front, $d_b$ is the distance to the vehicle behind, and $\eta$ is used to convert safety value to action, i.e. ego vehicle's acceleration.
The experiment results shown in Section \ref{sec:experiment} prove that the rule can achieve collision avoidance under different traffic densities.

\section{Loss-aware experience selection strategy}
\label{sec:loss-aware}

In the setting of the proposed IL in Section \ref{sec:FIL}, the experience, i.e. $(s,a)$ tuple, is generated by each vehicle and uploaded to ETs for training at $10Hz$. This will consume many communication resources when the number of vehicles near an intersection is enormous.
Inspired by \cite{DBLP:journals/corr/SchaulQAS15}, not all the experience, i.e. $(s_t,a_t,r_t,s_{t+1})$ tuple, is valuable for the RL model training.
This can also be analogous to IL.
This section introduces computing for communication, where extra computation is performed to reduce communication overhead.
In this section, the extra computation is placed on vehicles and edge nodes. Vehicles calculate the loss and compare the loss with thresholds given by edge nodes. Edge nodes produce a threshold for loss comparison.
Therefore, combined with the concept of computing for communication, a loss-aware experience selection strategy is proposed to throw away experience that helps to model training.

\begin{figure}[H]
  \centering
  \includegraphics[width=\linewidth]{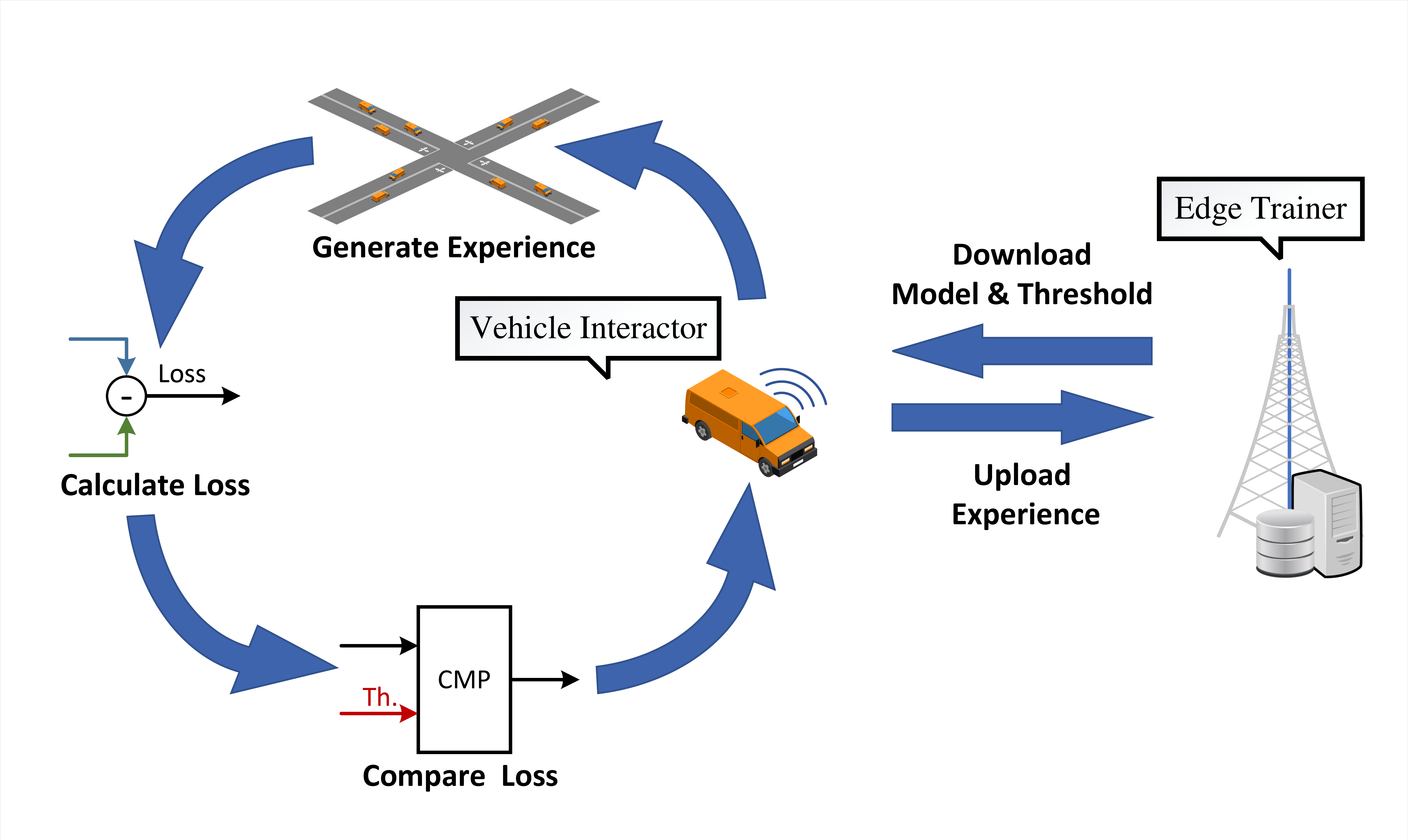}
  \caption{Loss-aware experience selection strategy.}
  \label{fig:Loss-aware-exp} %% label for entire figure
\end{figure}

As displayed in Fig.\ref{fig:Loss-aware-exp}, the proposed strategy is applied between vehicle interactors and edge trainers.
When a vehicle enters an intersection area, it requests edge trainers for the newest model and threshold.
Then, all vehicles interact with the environment using vehicular collision avoidance rules and generate experience $(s, a)$.
Then, the vehicle produces the action with the current model and state $s$, and output the action loss between rule and model.
The loss is compared with loss threshold $Th.$. If the action loss is larger than $Th.$, the corresponding experience $(s, a)$ is allowed to upload to the edge trainers, vice versa.
Finally, the received experiences support the edge trainers to perform the IL training and output a new threshold. The threshold is calculated on ETs as follows,
\begin{equation}
  \label{eq:th_cal}
  % Th. = ({L_{\max }} - {L_{\min }}) \times p + {L_{\min }},
  Th. = sort(\overrightarrow {Exp},"loss")[p \times B]["loss"]
\end{equation}
where $\overrightarrow {Exp}$ is a batch of experience for training, and its size is $B$. The function $sort(\cdot,"loss")$ sorts experiences according to the ascending order of loss value. $p$ is the discard rate. Eq.\ref{eq:th_cal} means that the threshold is the $(p \times B)^{th}$ smallest loss in the experience batch.
Because the experience will be partly discarded, the communication overhead will be saved.

\section{Experiment}
\label{sec:experiment}

\subsection{General setups}

The proposed DUCF algorithm is trained and evaluated in a self-designed intersection vehicle control platform, which is developed based on Python 3.5.
Each intersection consists of four driving directions, and vehicles, which are generated following the Poisson process with different densities, are allowed to go straight without steering.
The related vehicle control parameters are listed in Table.\ref{tab:Experiment_Parameter}.

\begin{table}
  \centering
  \caption{Experimental Parameters}
  \label{tab:Experiment_Parameter}
  \begin{tabular}[l]{@{}lc}
    \toprule
    \textbf{Parameter}               & \textbf{Value} \\
    \midrule

    \textbf{\emph{Simulator}}                         \\
    Lane length ($m$)                & $150$          \\
    Vehicle size ($m$)               & $2$            \\
    Velocity ($m/s$)                 & $[6,13]$       \\
    Initial velocity ($m/s$)         & $10$           \\
    Acceleration ($m/s^2$)           & $[-3,3]$       \\
    Discrete-time step $T$ ($s$)     & $0.1$          \\

    \midrule
    \textbf{\emph{Safety Value}}                      \\
    $\alpha_{s}$                     & $10$           \\
    $\beta_{s}$                      & $10$           \\
    $\alpha_{t}$                     & $1.5$          \\
    $\beta_{t}$                      & $2$            \\
    $\alpha_{acc}$                   & $1.5$          \\
    $\beta_{acc}$                    & $12$           \\
    $\lambda_{acc}$                  & $0.2$          \\
    $SV_{max}$                       & $20$           \\
    $SV_{min}$                       & $-20$          \\

    Conversion factor $\eta$         & $3$            \\
    Fusion factor $\omega$           & $0.2$          \\
    Weighting factor $\varepsilon $  & $0.5$          \\

    \midrule
    \textbf{\emph{Vehicle Selection}}                 \\
    Number of the closet vehicle $n$ & $5$            \\

    \bottomrule
  \end{tabular}
\end{table}

In the proposed schemes, the neural network (NN) is used to imitate the collision-free rule by minimizing the action loss between the NN and the rule. There is only one NN, containing three dense layers and two normalization layers, and ReLU is chosen as the activation function in hidden layers. The output layer is activated by $tanh(\cdot)$. To fit the range of acceleration, the output of the NN is multiplied by 3.
The complete hyper-parameters are listed in Table.\ref{tab:HyperParameter}.
However, due to poor interpretability and limited safety performance of end-to-end NN inference, a weighted operation is added as below,
\begin{equation}
  \label{eq:weighted_operation}
  a_{exe} = \varepsilon \times a_{NN} + (1-\varepsilon ) \times a_{rule},
\end{equation}
where $\varepsilon$ is a factor for smoothing the NN output $a_{NN}$ with the rule output $a_{rule}$ to ensure driving safety.
In the following experiment results, \emph{Model} denotes action output using NN only, \emph{Model+Rule} represents mixed output using NN and rule.

\begin{table}
  \centering
  \caption{Parameters for Neural Networks}
  \label{tab:HyperParameter}
  \begin{tabular}[l]{@{}lccccccc}
    \toprule
    % \makecell[c]{\textbf{Parameter}} & \multicolumn{4}{c}{\textbf{Value}}\\ 

    \textbf{Parameter}         & \textbf{Value}          \\
    \midrule
    % \midrule
    Discounted factor $\gamma$ & {0.8}                   \\
    Batch Size $B$             & {48}                    \\
    Soft update factor $\tau$  & {0.99}                  \\
    Episode                    & {50}                    \\
    Learning rate              & {0.001 $\rightarrow$ 0} \\
    Optimizer                  & {Adam}                  \\

    \midrule
    \textbf{\emph{Network Architecture}}                 \\

    Dense layer 1\#            & 64                      \\
    Dense layer 2\#            & 64                      \\
    Dense layer 3\#            & 1                       \\

    \bottomrule
  \end{tabular}
\end{table}

\subsection{Indicator}

To comprehensively evaluate the performance of the proposed vehicle control methods at intersections, three indicators are chosen, including safety, efficiency and discomfort.

Safety is the first indicator to evaluate the vehicle control. This paper treats the collision ratio $r_{collision}$ as the safety indicator, shown as below,
\begin{equation}
  \label{eq:metric_safe}
  {r_{collision}} = \frac{n_{collision}}{N_{veh}},
\end{equation}
where $n_{collision}$ is the number of collisions, and $N_{veh}$ is the total number of vehicles. A large $r_{collision}$ means the vehicle control algorithm is not capable of achieve collision avoidance. The algorithm in this article is designed to reduce the indicator value to 0.

While ensuring vehicles' safety, traffic efficiency is an indicator needs to be improved. This paper chooses $v_{avg}$ as the indicator, presented as below,
\begin{equation}
  \label{eq:metric_eff}
  {v_{avg}} = \frac{1}{{{N_{veh}}}}\sum\limits_{i = 1}^{{N_{veh}}} {\frac{{{l_{road}}}}{{{t_i}}}},
\end{equation}
where $t_i$ is the $i^{th}$ vehicle's travel time and $l_{road}$ is the length of the road, which is given in Table.\ref{tab:Experiment_Parameter}. A large $v_{avg}$ represents vehicles with the proposed algorithm driving faster, which also means a high throughput for the transportation system.

Driving discomfort is also an important indicator for vehicle control. As presented in \cite{DBLP:conf/cdc/ZhangMC17, katriniok2018distributed}, the driving discomfort can be defined as below,
\begin{equation}
  \label{eq:metric_comf}
  {J_{avg}} = \frac{1}{{{N_{veh}}}}\sum\limits_{i = 1}^{{N_{veh}}} {\sum\limits_t {j_{i,t}^2} } ,
\end{equation}
where $j_{i,t}$ is the $i^{th}$ vehicle's jerk. The jerk is defined by ${j_{i,t}} = {\dot a_{i,t}}$, and $a_{i,t}$ is the $i^{th}$ vehicle's acceleration at time step $t$.
A large $J_{avg}$ indicates more frequent or sharp acceleration and deceleration, resulting in more severe driving discomfort.

\subsection{Results Analysis}

The whole experiment contains three parts to evaluate the corresponding schemes.
In the first part, the proposed IL scheme is compared with an RL algorithm and pure rule, described in Section \ref{sec:rule} in terms of training, and three indicators mentioned above.
Secondly, the proposed density-aware federated learning algorithm is compared to the local training with different traffic densities.
Finally, to verify the effectiveness of the proposed loss-aware experience selection strategy, the three indicators and communication reduction are treated as metrics to compare the strategy with different $p$.

\begin{figure*}
  \subfigure[Imitation Learning-Safety]{
    \label{fig:IL_safety}
    \includegraphics[width=0.31\textwidth]{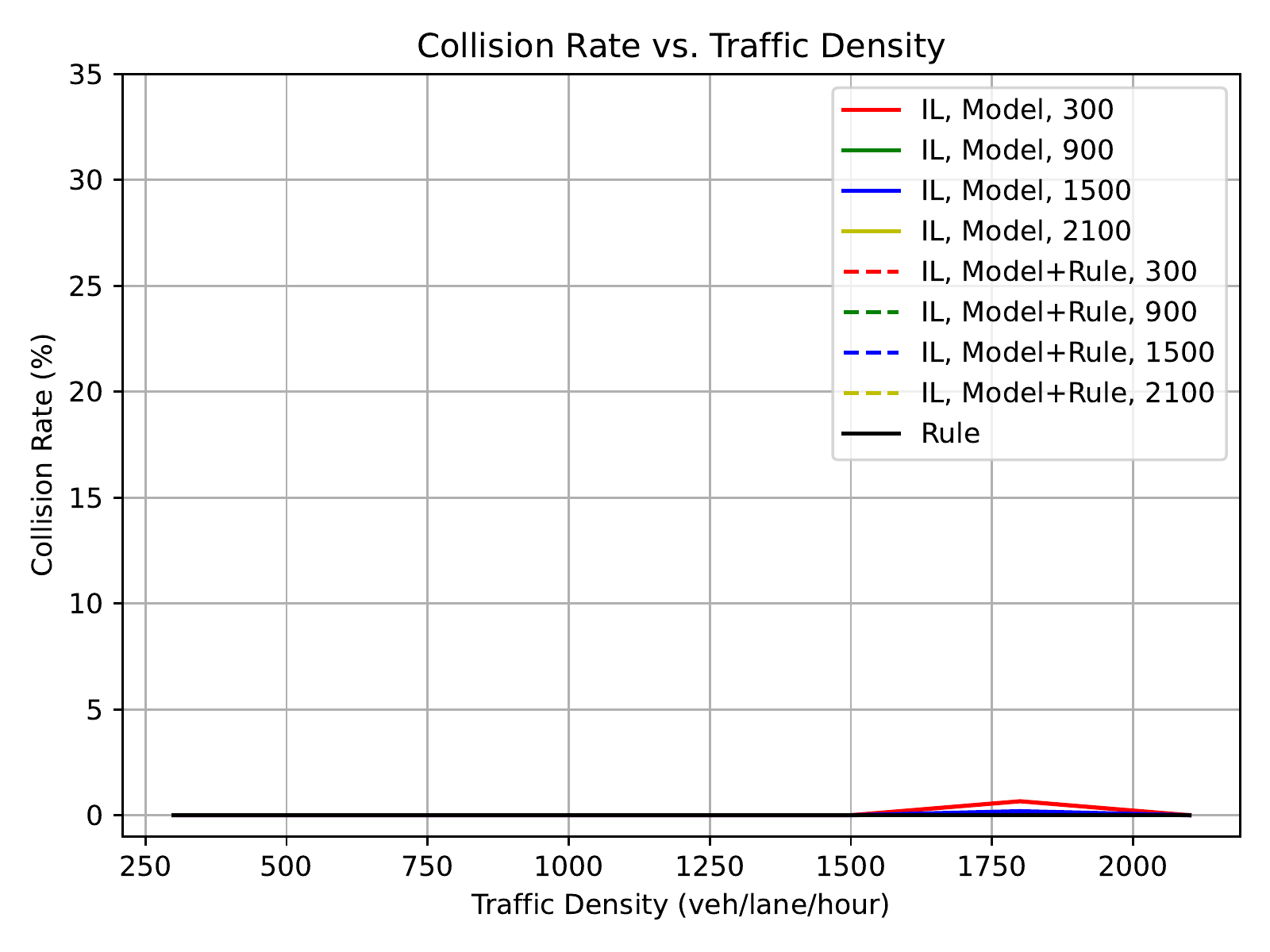}
  }
  \subfigure[Imitation Learning-Discomfort]{
    \label{fig:IL_discomf}
    \includegraphics[width=0.31\textwidth]{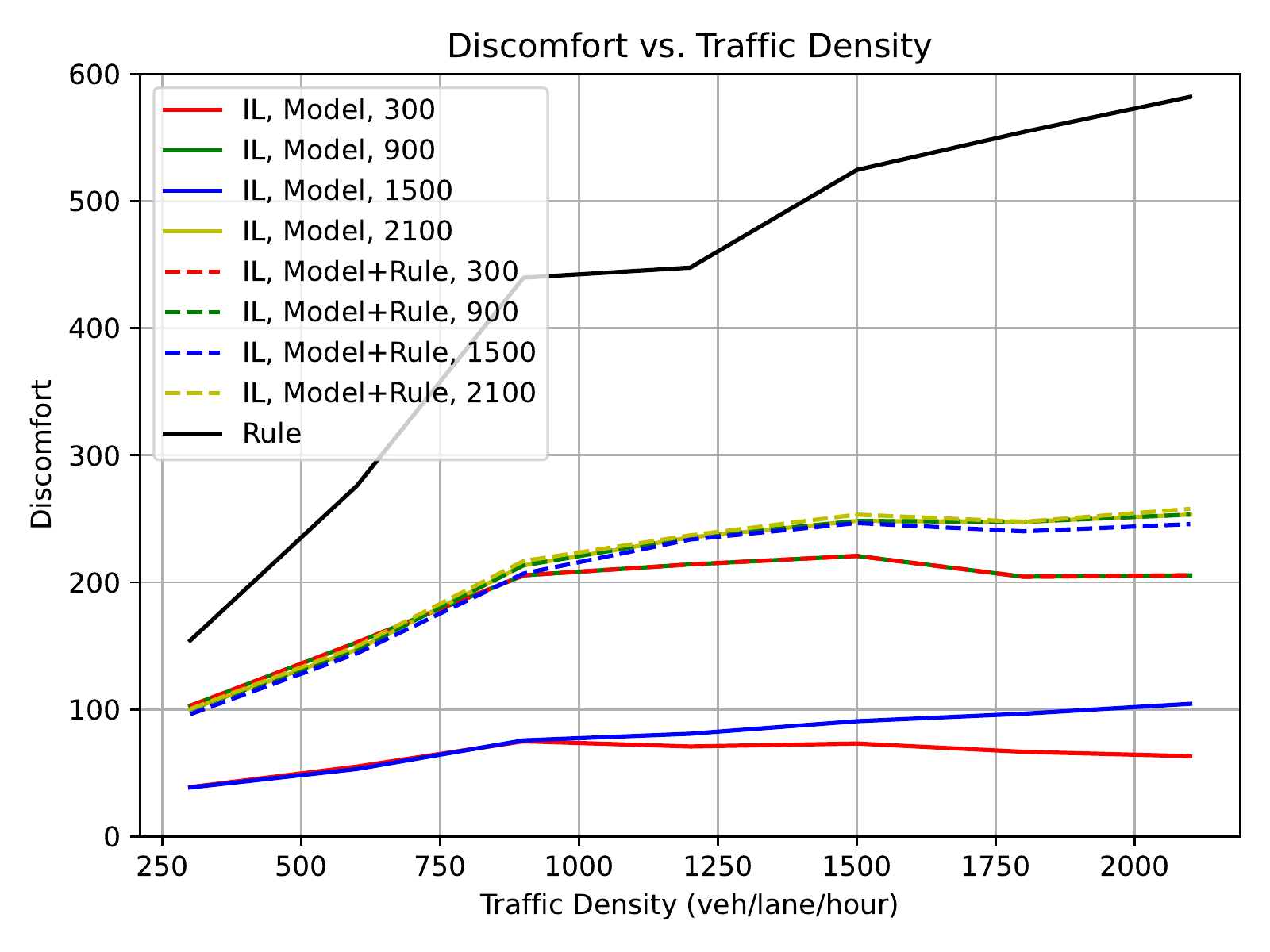}
  }
  \subfigure[Imitation Learning-Efficiency]{
    \label{fig:IL_velocity}
    \includegraphics[width=0.31\textwidth]{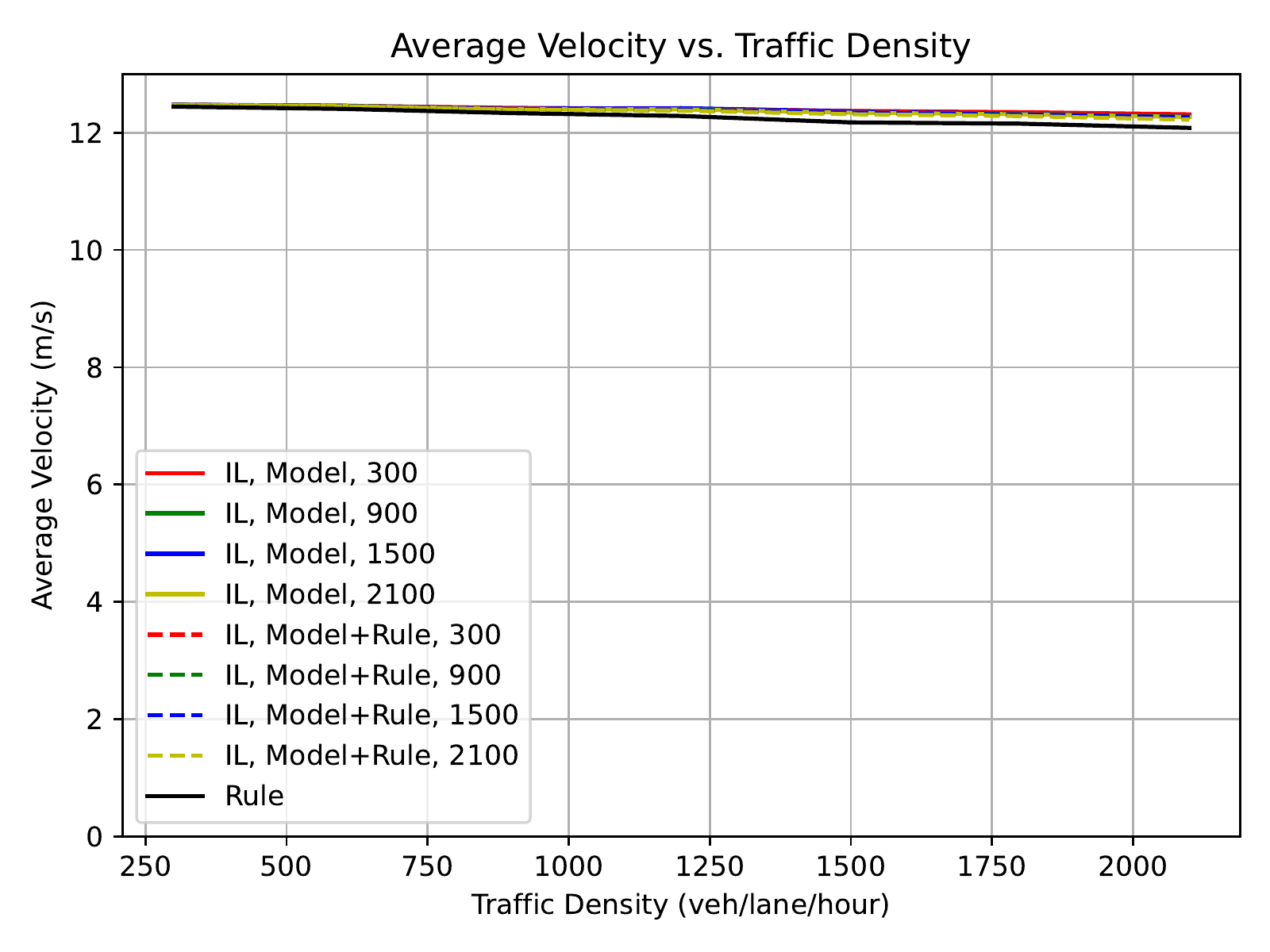}
  }
  \caption{The performance of the proposed imitation learning.}
  \label{fig:IL_perf}
\end{figure*}
% =============================================================================
\begin{figure*}
  \subfigure[Reinforcement Learning-Safety]{
    \label{fig:RL_safety}
    \includegraphics[width=0.31\textwidth]{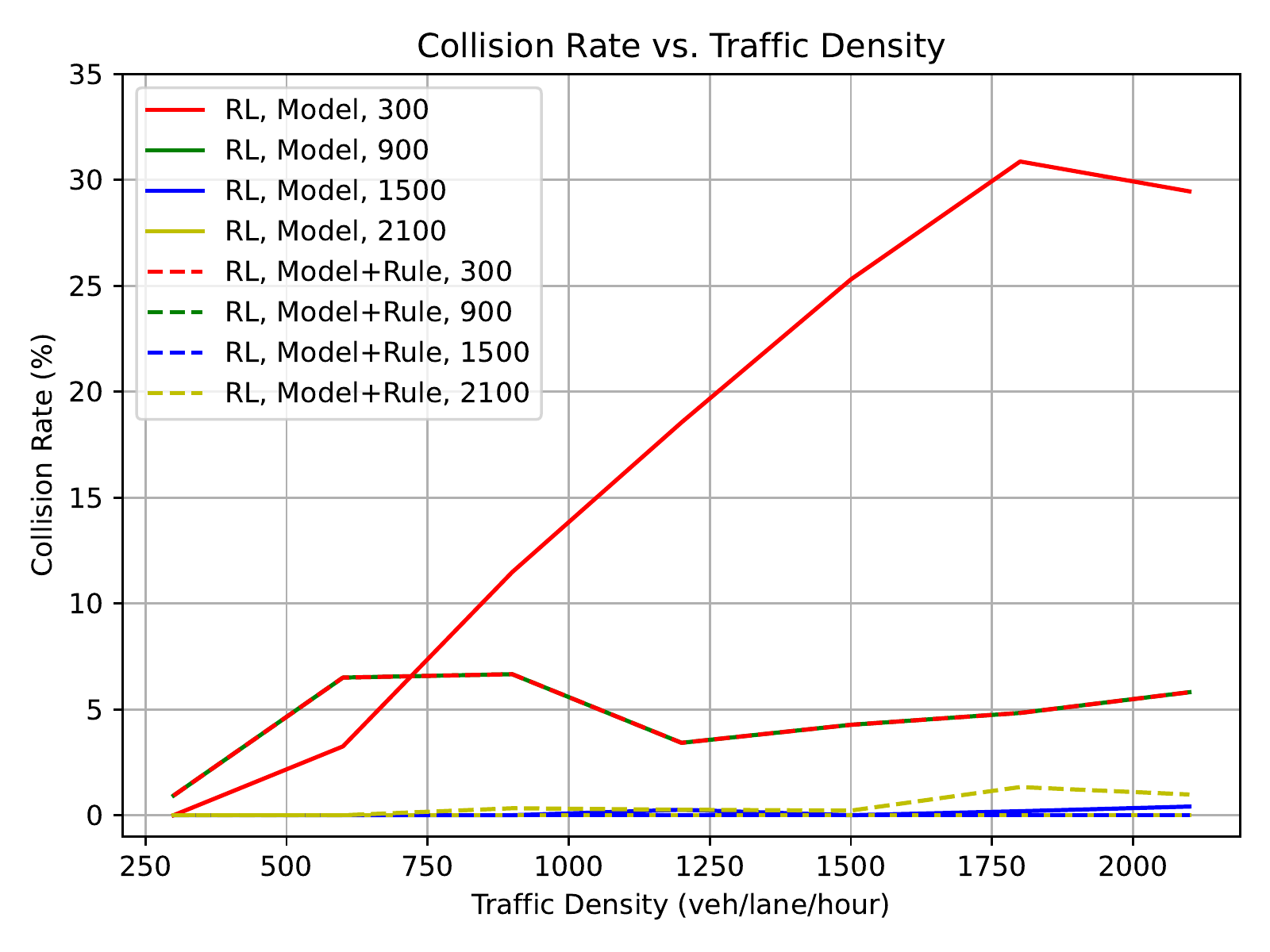}
  }
  \subfigure[Reinforcement Learning-Discomfort]{
    \label{fig:RL_discomf}
    \includegraphics[width=0.31\textwidth]{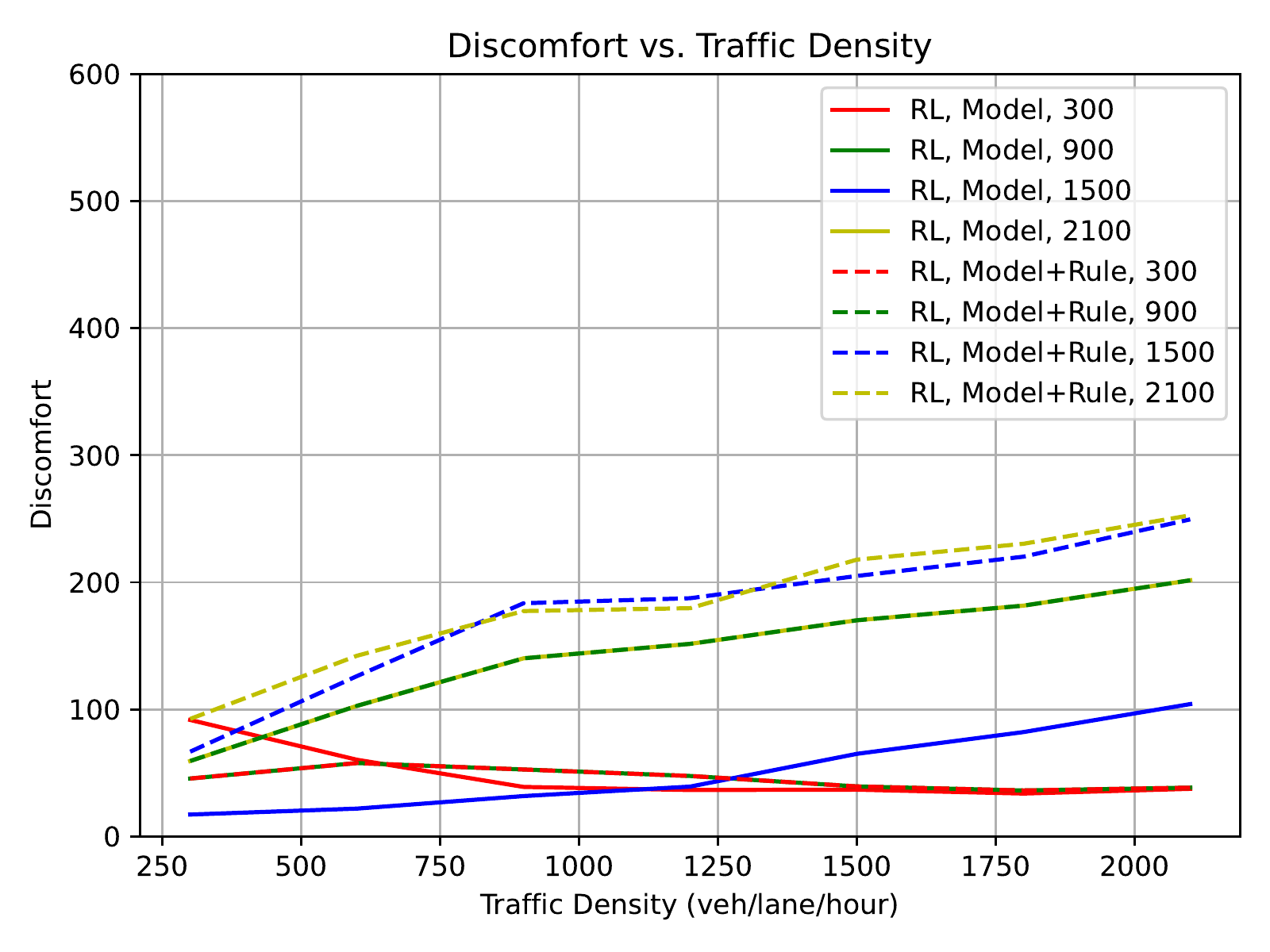}
  }
  \subfigure[Reinforcement Learning-Efficiency]{
    \label{fig:RL_velocity}
    \includegraphics[width=0.31\textwidth]{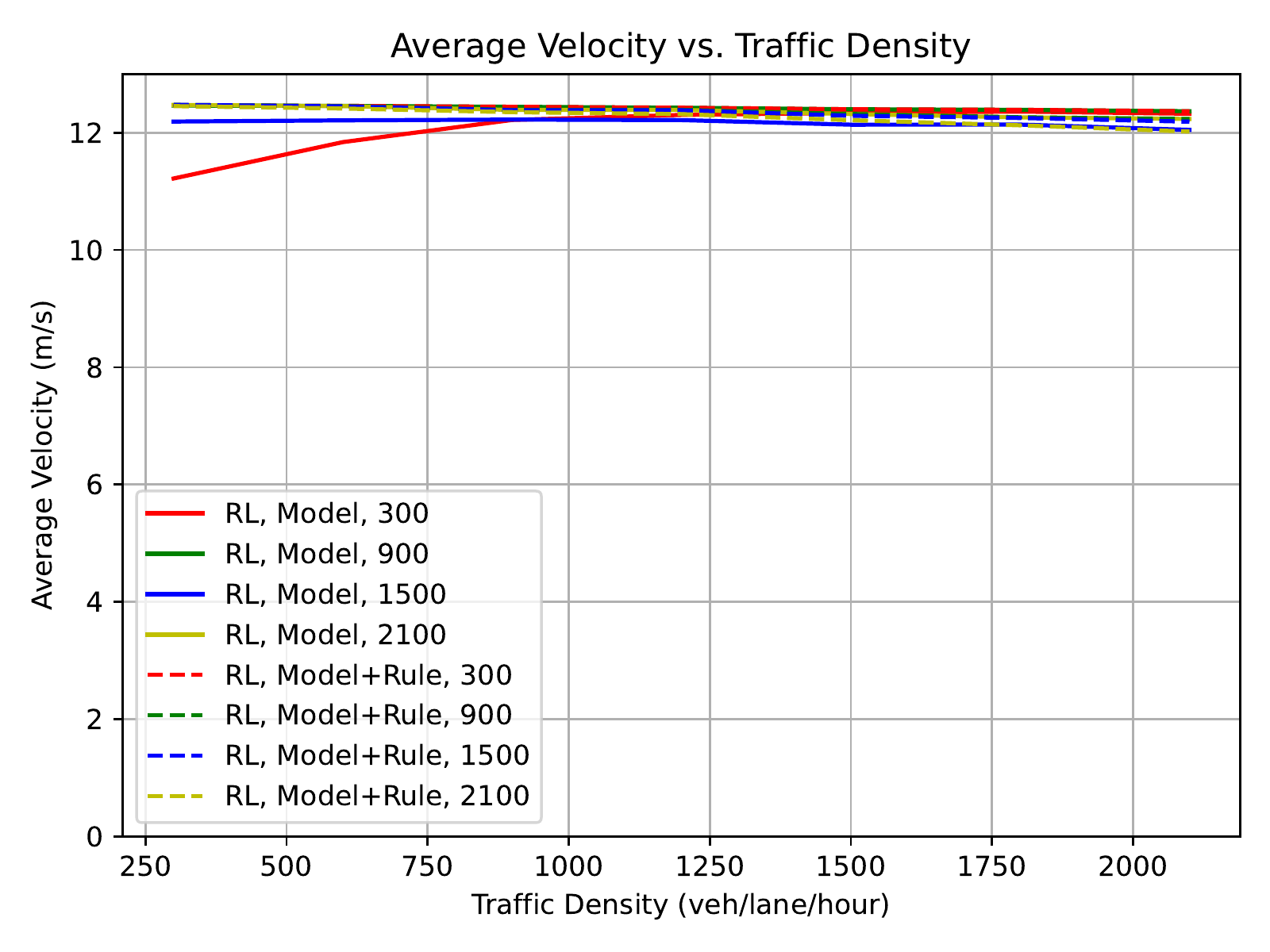}
  }
  \caption{The performance of the benchmark reinforcement learning.}
  \label{fig:RL_perf}
\end{figure*}
% =============================================================================
\begin{figure*}
  \subfigure[Federated Learning-Safety]{
    \label{fig:FL_safety}
    \includegraphics[width=0.31\textwidth]{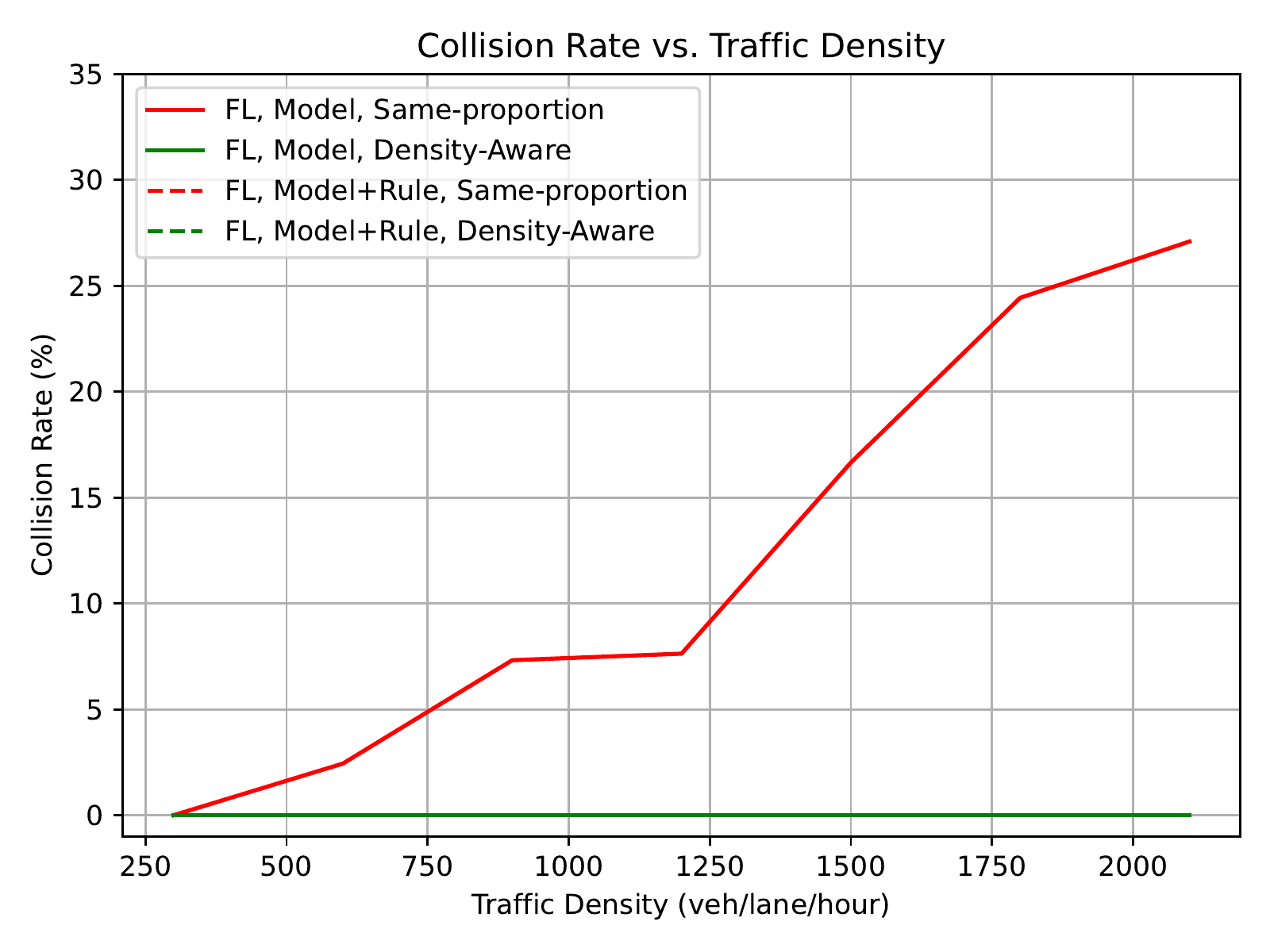}
  }
  \subfigure[Federated Learning-Discomfort]{
    \label{fig:FL_discomf}
    \includegraphics[width=0.31\textwidth]{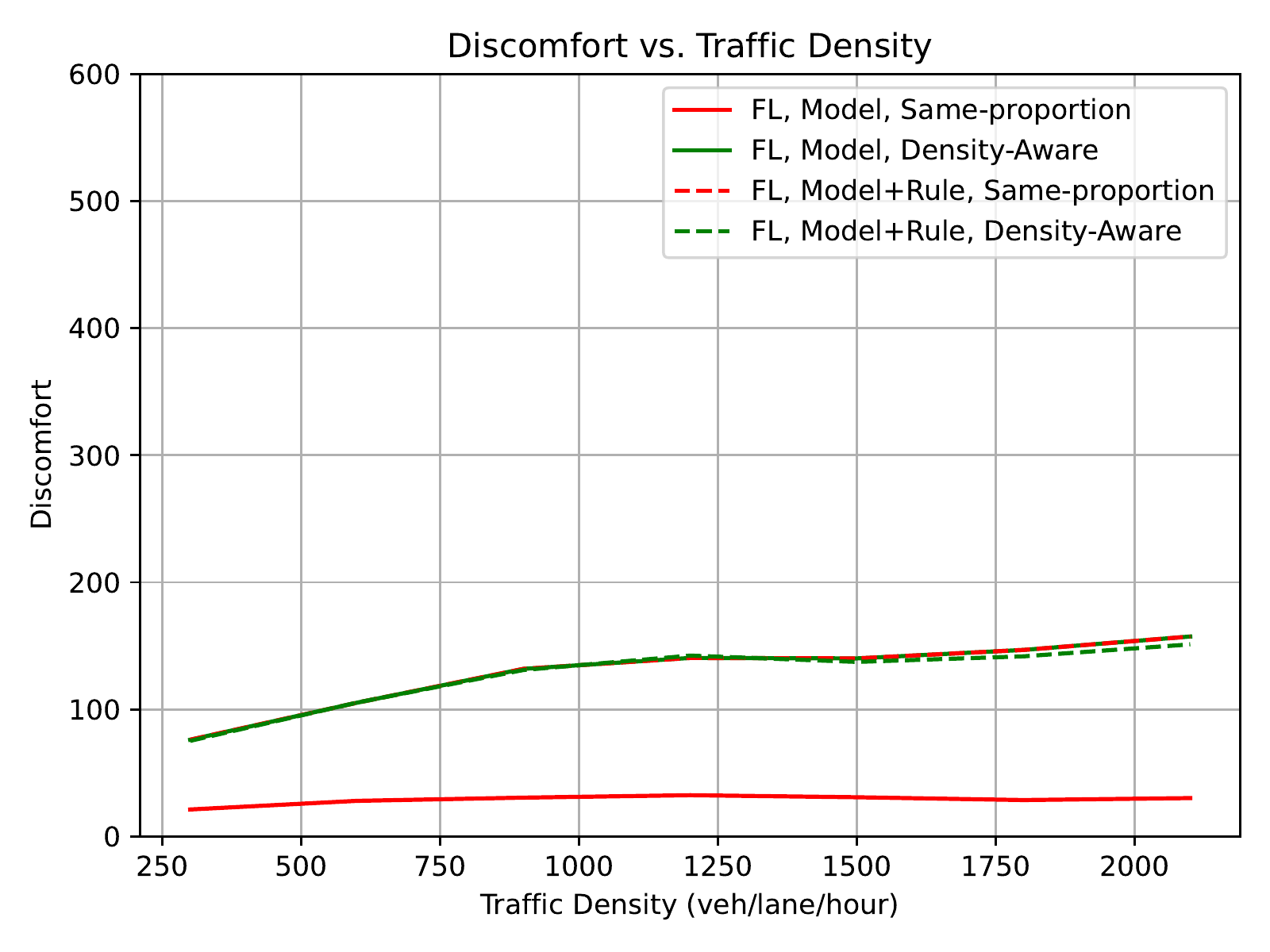}
  }
  \subfigure[Federated Learning-Efficiency]{
    \label{fig:FL_velocity}
    \includegraphics[width=0.31\textwidth]{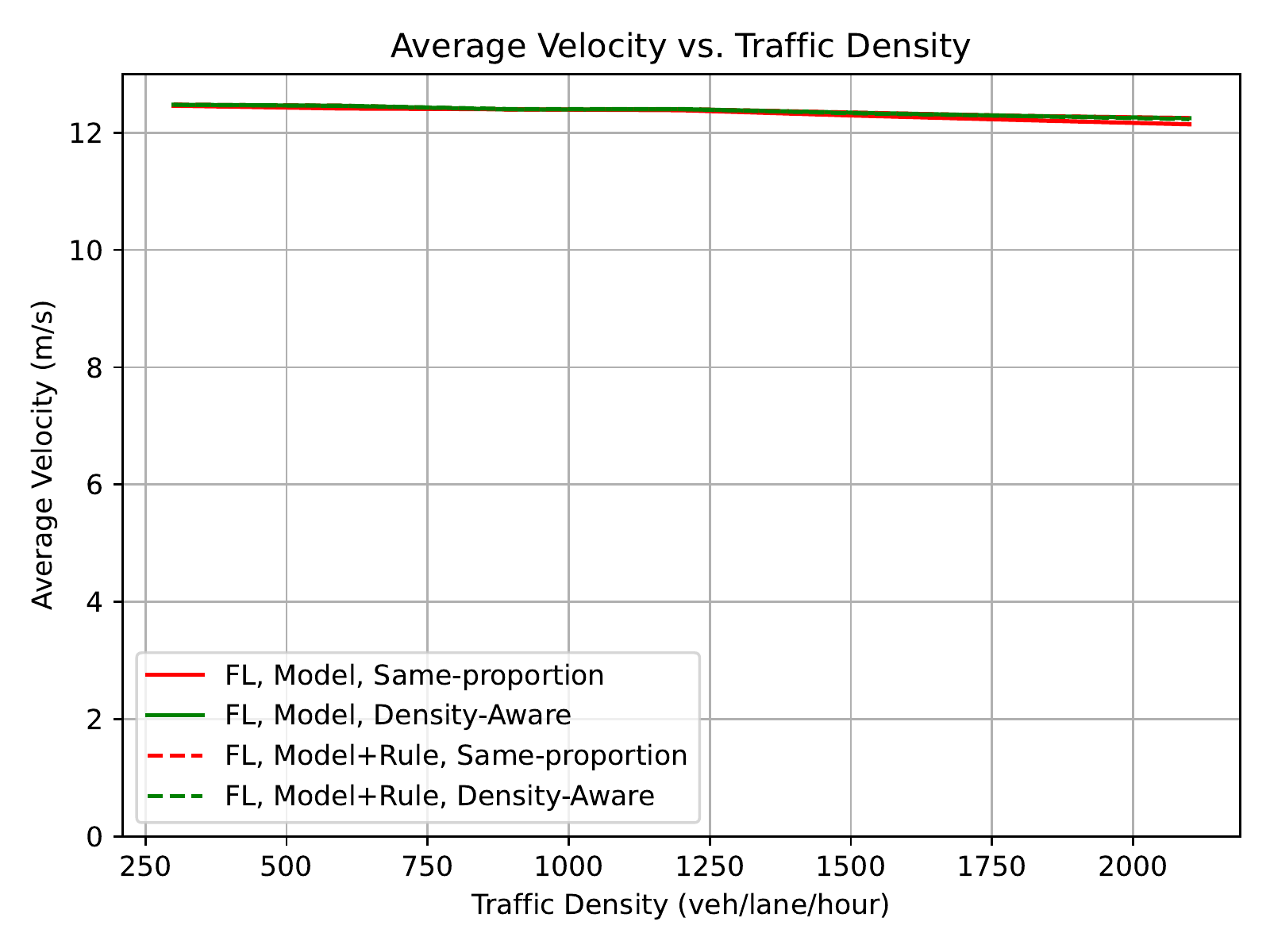}
  }
  \caption{The performance of the proposed federated learning.}
  \label{fig:FL_perf}
\end{figure*}

Fig.\ref{fig:IL_perf} depicts the performance of the proposed IL scheme for varying traffic densities from 300 to 2100 veh/lane/hour. The IL scheme is trained in four traffic densities (i.e. 300/900/1500/2100 veh/lane/hour).
The black lines in these sub-figures represent the set of collision avoidance rules described in Section \ref{sec:rule}, and the results show that the rules can achieve good collision avoidance and traffic efficiency but poor driving comfort. This is because the safety value emphasized in this set of rules is to evaluate the surrounding danger degree according to space, time and acceleration, so as to avoid the collision, but it does not suppress the change of acceleration.
In addition, the results also demonstrate that the proposed IL algorithm can help the model learn the ability of collision avoidance and efficiency improvement from the rules.
Furthermore, only the model trained in low-density produces a small ratio of collisions in a high-density evaluation environment. This is because, in the low-density scene, the experience samples are concentrated in large inter-vehicle distances. Such sample distribution contributes to insufficient model training, which causes the model unable to cope with the high-density (i.e. small inter-vehicle distance) scenarios.
Moreover, compared with the rule, the policy obtained from IL can reduce the discomfort by 55.71\%. It can be explained that rule-making inevitably introduces threshold trigger mode, and the state that has not yet triggered the threshold can not be fully mapped to the action. The model training helps complete the state-action mapping by gradually approaching the rule with a low learning rate.

Fig.\ref{fig:RL_perf} shows the performance of a benchmark RL algorithm for different traffic densities. In the comparison of Fig.\ref{fig:IL_perf} and Fig.\ref{fig:RL_perf}, it can find that IL outperforms RL in terms of safety and efficiency.
Obviously, RL is more dependent on samples, which can be easily found from the low-density training and high-density evaluation in Fig.\ref{fig:RL_safety}.
Using the weighted operation mentioned in Eq.\ref{eq:weighted_operation}, rules can contribute to IL model achieving collision-free in all cases, but they can not effectively assist RL model.

Fig.\ref{fig:FL_perf} presents the impact of the model aggregation operation in DUCF.
Two aggregation methods (Same-proportion and Density-Aware) are evaluated.
Our training scenario uses four traffic densities (300/900/1500/2100 veh/lane/hour).
The Same-proportion means the four models have the same weight, and density-aware means the weight is $1:3:5:7$. The density-aware method outperforms same-proportion in all three indicators.
In addition, because the global model aggregates the model parameters under different densities, the discomfort is further reduced by 41.37\% compared with the trained policy under any single traffic density.
Taking a comprehensive view of Fig.\ref{fig:IL_perf}-\ref{fig:FL_perf}, the average speed is basically unchanged, and there is a conversion relationship between collision rate and discomfort.
Models with higher collision rates are more conservative, but lead to better comfort and vice versa.

\begin{figure}
  \centering
  \includegraphics[width=\linewidth]{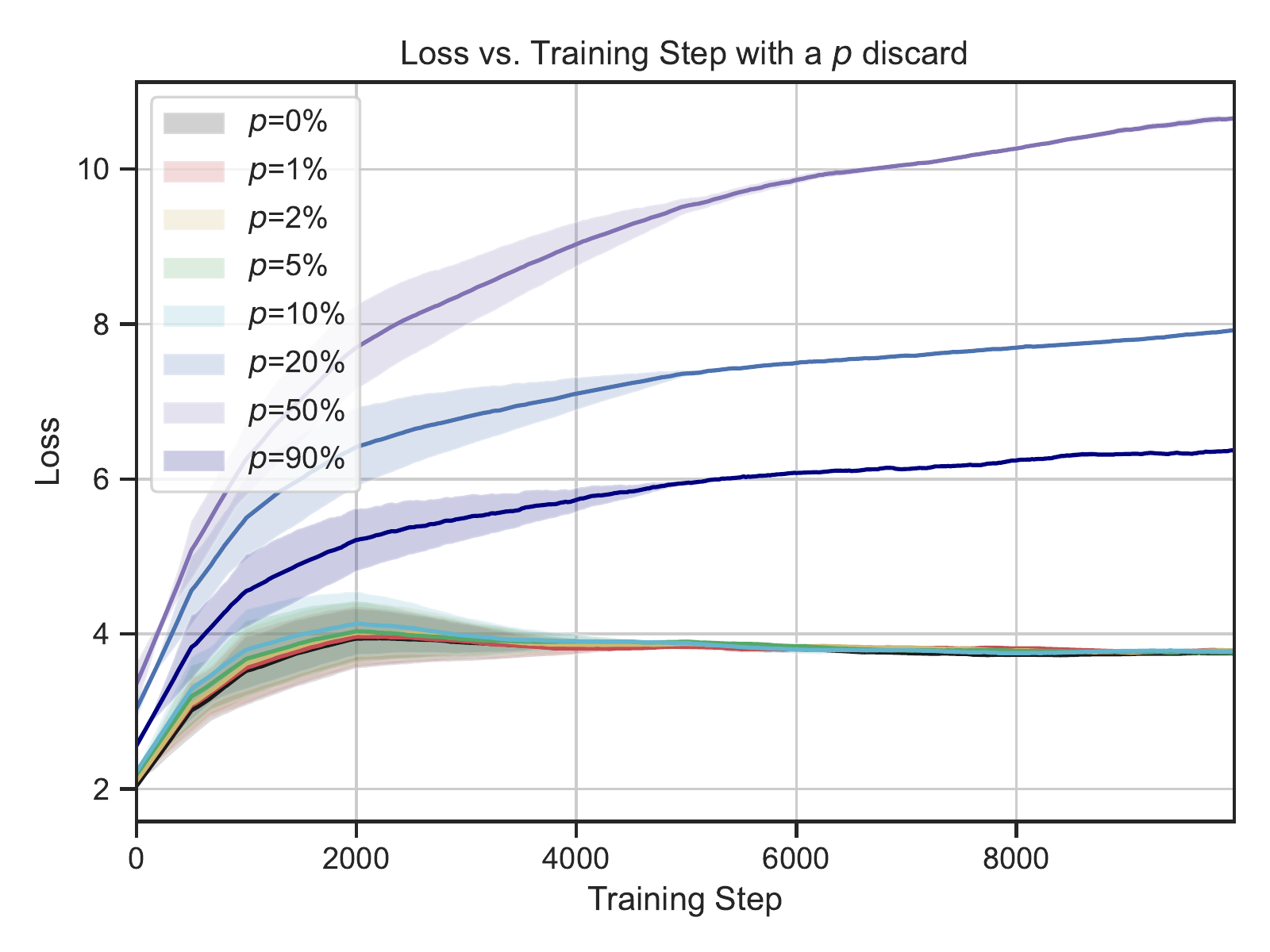}
  \caption{The loss value with different discard factor $p$.}
  \label{fig:save_loss} %% label for entire figure
\end{figure}

\begin{figure}
  \centering
  \includegraphics[width=\linewidth]{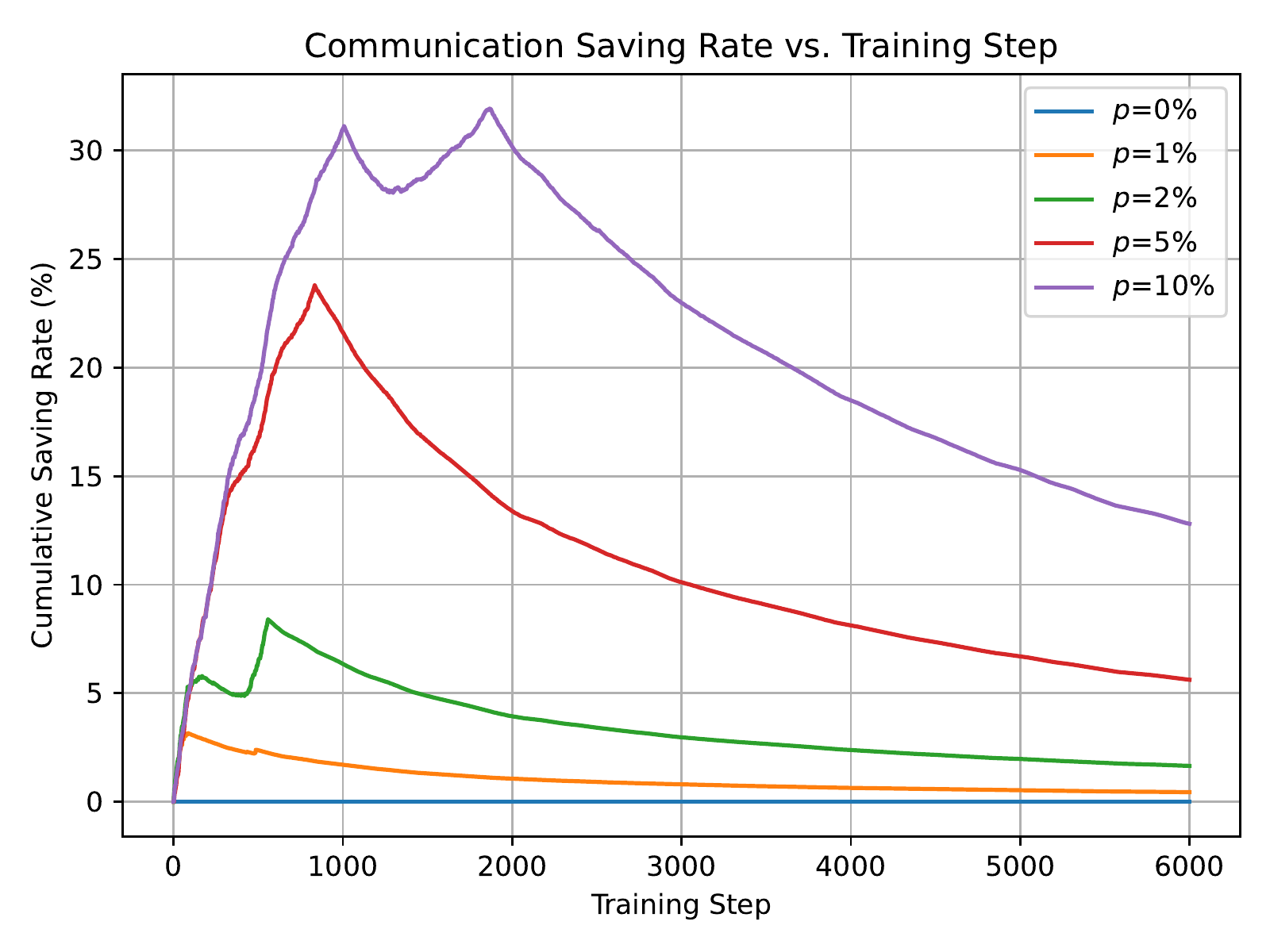}
  \caption{Cumulative communication savings with different discard factor $p$.}
  \label{fig:save_ratio} %% label for entire figure
\end{figure}

Fig.\ref{fig:save_loss} depicts the loss curves with different discard factor $p$.
When $p$ is not more than 10\% With the training steps, the trained models can be convergent and stable.
The results demonstrate that the IL model is more difficult to converge as the number of discarded experiences increases. But the IL model can still converge when the number of discarded experiences are limited within a certain range.
In the following analysis, only the convergent model (i.e. models with $p = 1\%, 2\%, 5\%, 10\%$) is consider.
In Fig.\ref{fig:save_ratio}, the cumulative communication savings is presented.  According to Fig.\ref{fig:save_loss}, the models converge in 6000 steps, so only the communication overhead before 6000 steps is counted.A higher $p$ brings less communication cost. When the model converges, the communication overhead can be saved 0.44\%, 1.65\%, 5.6\% and 12.80\%, respectively.
From the performance shown in Table.\ref{tab:save_perf}, the models are more conservative, that is, higher collision rate and comfort.

\begin{table}[]
  \caption{IL Performance with discard factors}
  \label{tab:save_perf}
  \begin{tabular}{llcccc}
    \toprule
    \multicolumn{2}{l}{Discard factor}              & 1\%        & 2\%    & 5\%    & 10\%            \\ \midrule
    \multirow{2}{*}{\begin{tabular}[c]{@{}l@{}}Collision\\ Rate\end{tabular}} & Model      & 36\%   & 39\%   & 44\%   & 42\%   \\
                                                    & Model+Rule & 0\%    & 0\%    & 0\%    & 0\%    \\ \midrule
    \multirow{2}{*}{Discomfort}                     & Model      & 21.55  & 13.08  & 11.11  & 23.85  \\
                                                    & Model+Rule & 137.15 & 113.23 & 108.83 & 130.83 \\ \midrule
    \multirow{2}{*}{\begin{tabular}[c]{@{}l@{}}Average\\ Velocity\end{tabular}} & Model      & 12.22  & 12.06  & 12.06  & 12.24  \\
                                                    & Model+Rule & 12.21  & 12.21  & 12.19  & 12.26  \\ \bottomrule
  \end{tabular}
\end{table}

\section{Conclusion}
\label{sec:conclusion}
This paper has proposed a FL-based vehicle control framework to address the problem of raw data interaction limitation. The framework contains three parts, interactors, trainers and an aggregator.
Under the framework, a density-aware model aggregation is proposed for intersections with different traffic densities.
Then, an IL algorithm is proposed for action cloning from a set of collision avoidance rules to improve the safety capability of end-to-end learning.
Furthermore, a loss-aware experience selection strategy is explored to reduce communication overhead via additional computation on interactors and trainers.
The extensive experiment reveals that the proposed IL algorithm obtains the ability to avoid collisions and reduces discomfort by 55.71\%. The density-aware model aggregation in FL framework can further reduce discomfort by 41.37\%, and the experience selection scheme can reduce the communication overhead by 12.80\% while ensuring convergence.

Our main future work will be focused on the modeling and theoretical analysis of the relationship between the interactors and trainers in terms of communication and model training. Based on this analysis, we believe that it will help speed up the model training and significantly reduce the communication overhead.

% or


% \appendices

% \section{}

\section*{Acknowledgment}

The authors would like to thank...

\ifCLASSOPTIONcaptionsoff
  \newpage
\fi



%\bibliographystyle{IEEEtran}
\bibliographystyle{IEEEtran}
\bibliography{IEEEabrv,ref}



% \end{thebibliography}


% \end{IEEEbiography}


%\newpage

\begin{IEEEbiography}[{\includegraphics[width=1in,height=1.25in,clip,keepaspectratio]{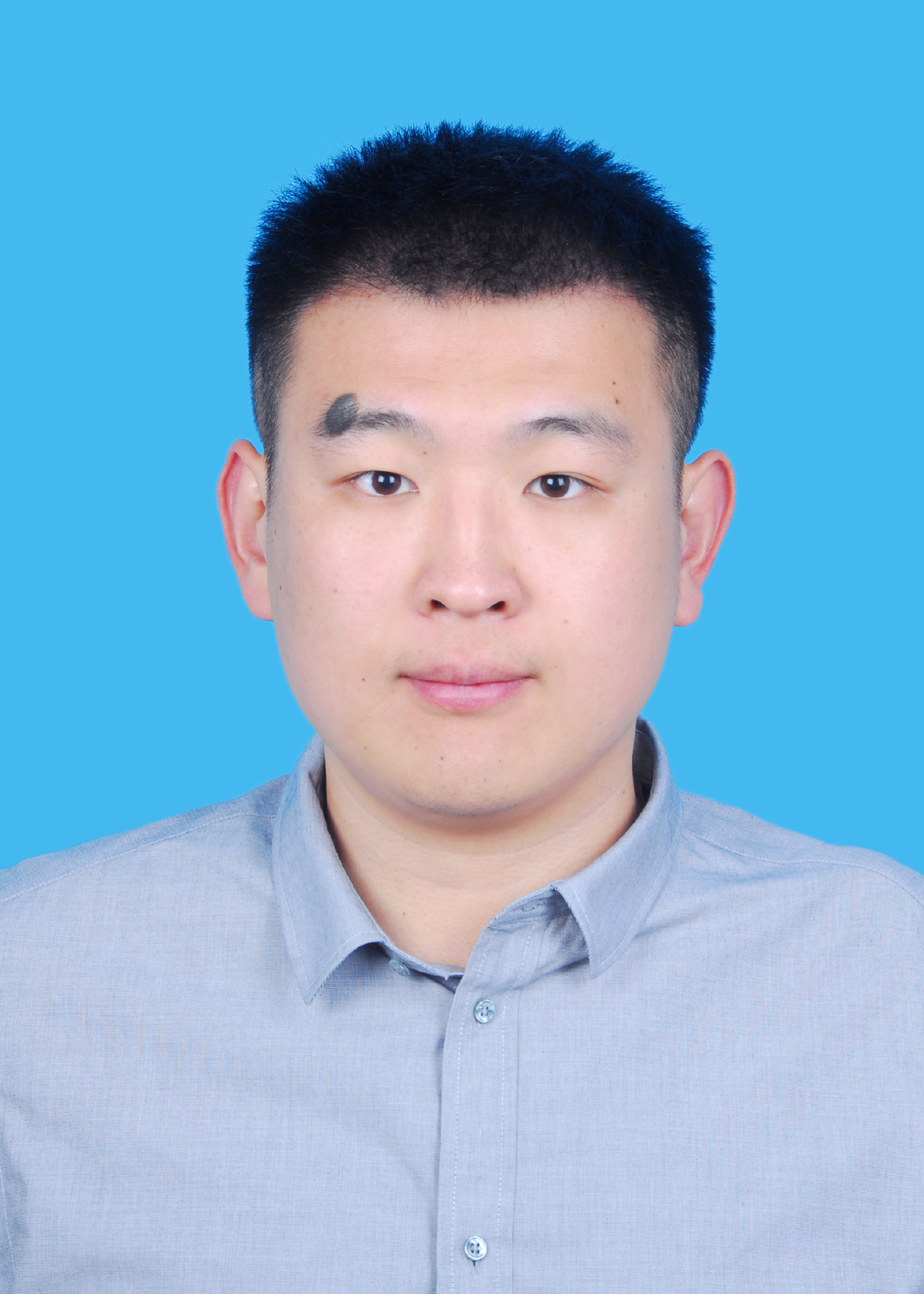}}]{Tianhao Wu}
  received the B.E. degree from Harbin Institute of Technology (HIT), China, in 2015. He is pursuing a PhD degree in Information and Communication Engineering, Beijing University of Posts and Telecommunications (BUPT), China. His research interests include cooperative intelligent transportation systems and multi-agent systems.
\end{IEEEbiography}

\begin{IEEEbiography}[{\includegraphics[width=1in,height=1.25in,clip,keepaspectratio]{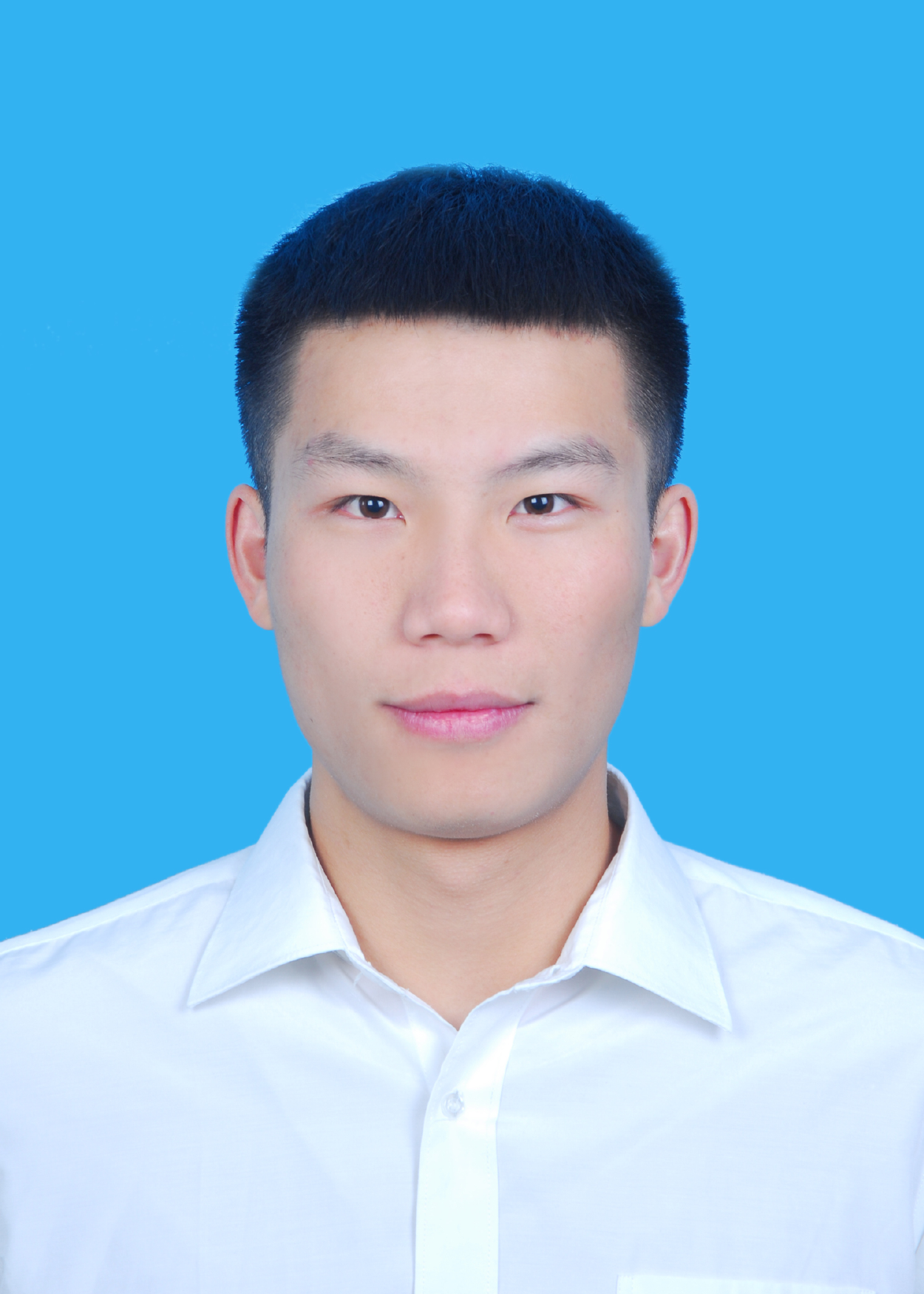}}]{Mingzhi Jiang}
  received the B.E. degree in electronics information engineering from YanShan University (YSU), China, in 2019. He is currently pursuing the M.S. degree with the School of Artificial Intelligence, Beijing University of Posts and Telecommunications (BUPT), Beijing, China. His current work focuses on the optimization of collaborative intelligent transportation systems and the application of artificial intelligence. His research interests include edge computing, deep learning, and reinforcement learning.
\end{IEEEbiography}

\begin{IEEEbiography}[{\includegraphics[width=1in,height=1.25in,clip,keepaspectratio]{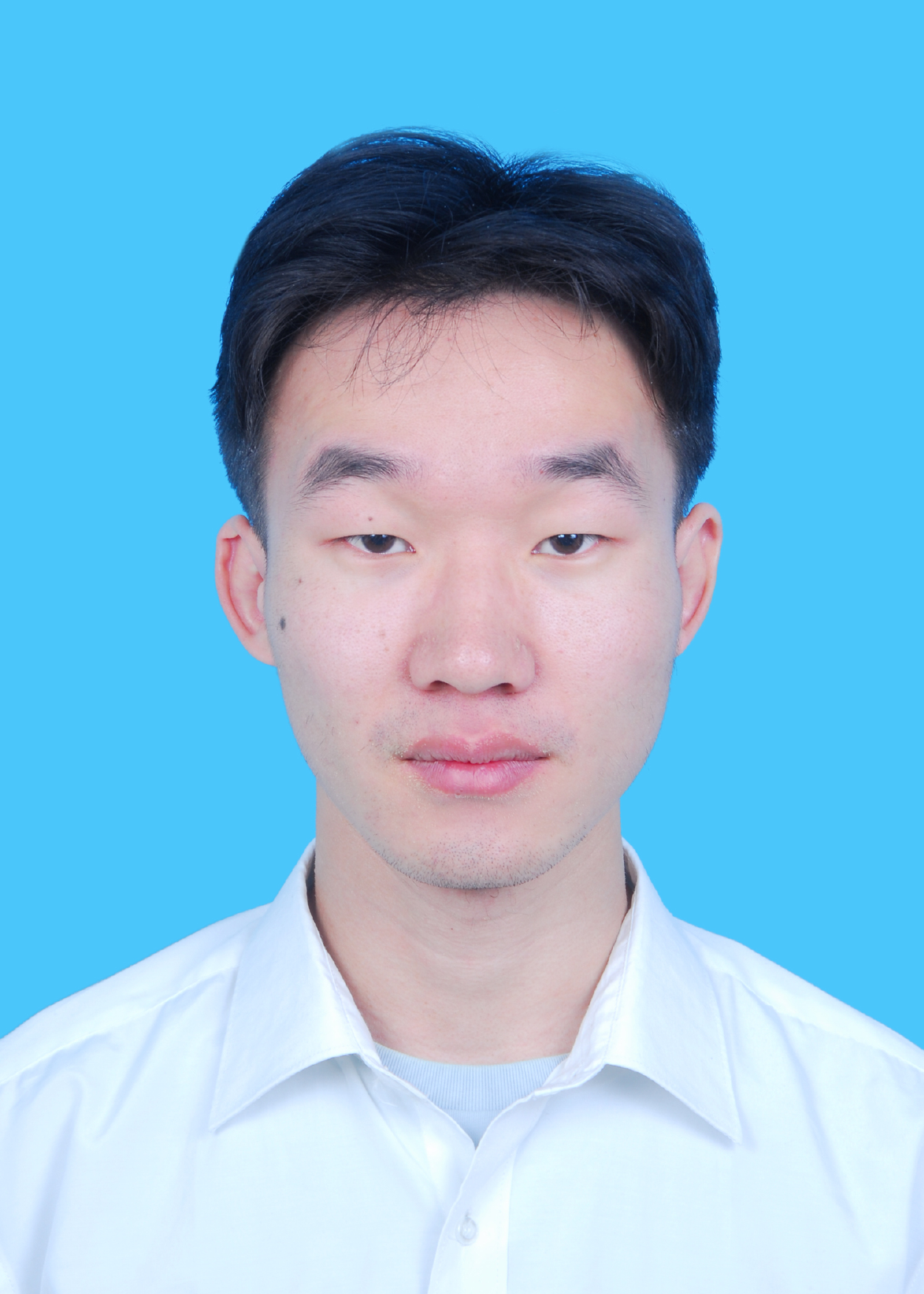}}]{Yinhui Han}
  received the B.E. degree in electronic science and technology from Xidian University (XDU), Xi’an, China, in 2019. He is currently pursuing an M.S. degree from the School of Artificial Intelligence, Beijing University of Posts and Telecommunications (BUPT), Beijing, China. His research interests are self-driving simulation platform and path planning.
\end{IEEEbiography}

\begin{IEEEbiography}[{\includegraphics[width=1in,height=1.25in,clip,keepaspectratio]{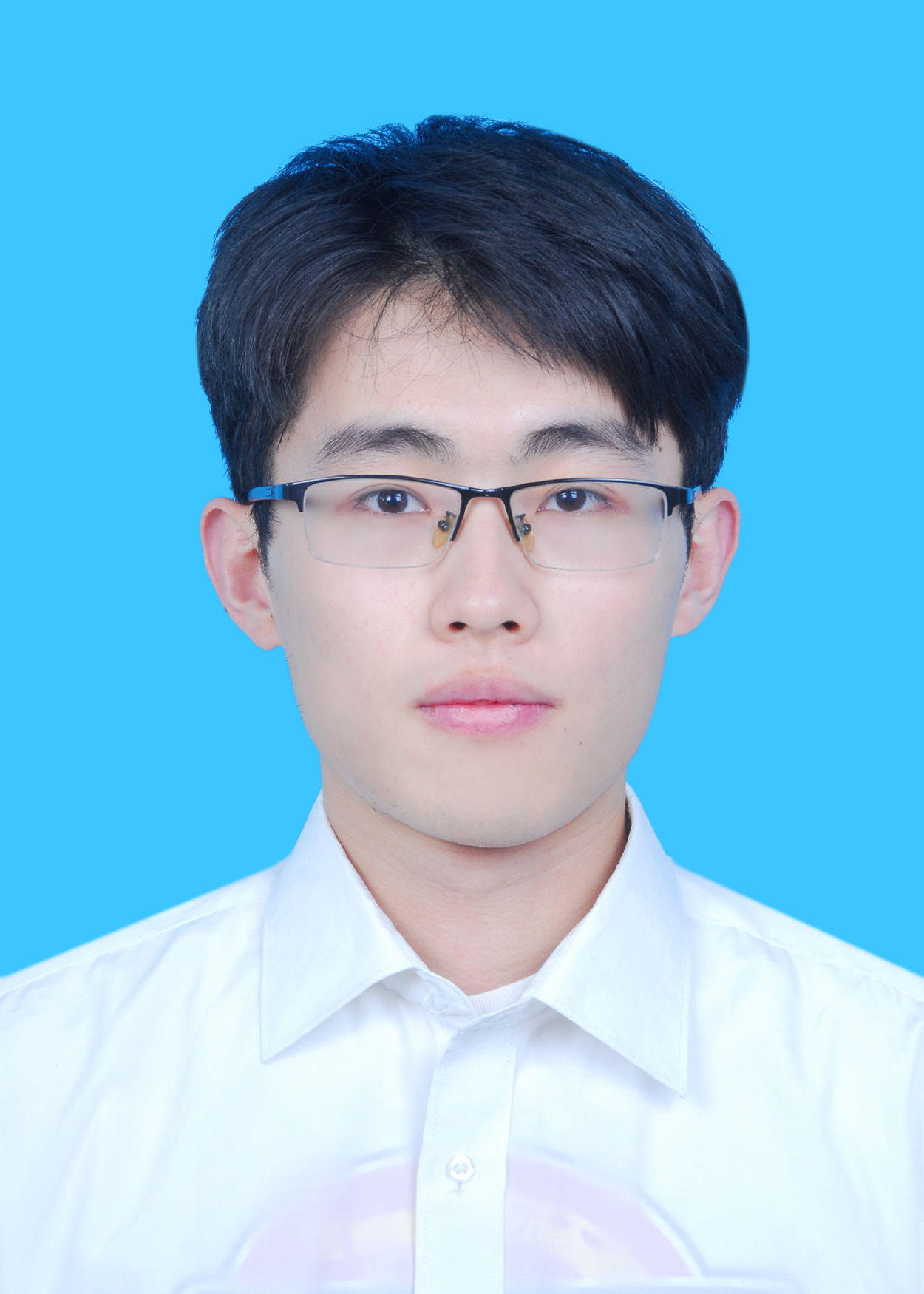}}]{Zheng Yuan}
  received the B.E. degree in information engineering from Beijing University of Posts and Telecommunications (BUPT), Beijing, China, in 2021. He is currently pursuing an M.S. degree from the School of Artificial Intelligence, Beijing University of Posts and Telecommunications (BUPT), Beijing, China. His research interests are self-driving and reinforcement learning.
\end{IEEEbiography}

% \begin{IEEEbiography}[{\includegraphics[width=1in,height=1.25in,clip,keepaspectratio]{author_ZheWang.pdf}}]{Zhe Wang}
%   received the B.E. degree in electronics and information engineering from Xi’an University of Posts and Telecommunications (XUPT), Xi’an, China, in 2019. He is currently pursuing the M.S. degree with the School of Artificial Intelligence, Beijing University of Posts and Telecommunications (BUPT), Beijing, China. His research interests are operation research in traffic congestion management, traffic assignment, and signal timing optimization.
% \end{IEEEbiography}

% \begin{IEEEbiography}[{\includegraphics[width=1in,height=1.25in,clip,keepaspectratio]{author_XinhangLi.pdf}}]{Xinhang Li}
%   received the B.E. degree in commuincation engineering from Beijing University of Posts and Telecommunications (BUPT), Beijing, China, in 2021. He is currently pursuing the Ph.D degree with the School of Artificial Intelligence, Beijing University of Posts and Telecommunications (BUPT), Beijing, China. His research interests is joint optimization of self-driving and communication.
% \end{IEEEbiography}

\begin{IEEEbiography}[{\includegraphics[width=1in,height=1.25in,clip,keepaspectratio]{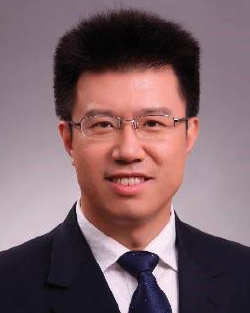}}]{Lin Zhang}
  received the B.S. and Ph.D. degrees from the Beijing University of Posts and Telecommunications (BUPT), Beijing, China, in 1996 and 2001, respectively. He is the vice chancellor of Beijing Information Science and Technology University (BISTU). He was a Postdoctoral Researcher with the Information and Communications University, Daejeon, Korea, from December 2000 to December 2002. He went to Singapore and held a Research Fellow position with Nanyang Technological University, Singapore, from January 2003 to June 2004. He joined BUPT in 2004 as a Lecturer, then an Associate Professor in 2005, and a Professor in 2011. He served the BUPT as the Director of Faculty Development Center, the Deputy Dean of Graduate School and the Dean of School of Information and Communication Engineering. He has authored more than 120 papers in referenced journals and international conferences. His research interests include mobile cloud computing and Internet of Things.
\end{IEEEbiography}


%\vfill


\end{document}